\documentclass[a4paper,fleqn,usenatbib]{mnras}

\usepackage{newtxtext,newtxmath}

\usepackage[T1]{fontenc}
\usepackage{ae,aecompl}

\usepackage{graphicx}	
\usepackage{amsmath, bm}	
\usepackage{amssymb}	

\title[EM heat flux instabilities]{Clarifying the solar wind heat-flux instabilities}

\author[S.M.Shaaban et al.]{
S. M. Shaaban,$^{1,2}$\thanks{E-mail: shaaban.mohammed@kuleuven}
M. Lazar,$^{1,3}$
S. Poedts$^{1}$
\\
 $^{1}$Centre for Mathematical Plasma Astrophysics, KU Leuven, Celestijnenlaan 200B, B-3001 Leuven, Belgium.\\
$^{2}$Theoretical Physics Research Group, Physics Department, Faculty of Science, Mansoura University, 35516, Mansoura, Egypt.\\
$^{3}$Institut f\"ur Theoretische Physik, Lehrstuhl IV: Weltraum- und Astrophysik, Ruhr-Universit\"at Bochum, D-44780 Bochum, Germany. 
}

\date{Accepted XXX. Received YYY; in original form ZZZ}

\pubyear{2018}

\begin{document}
\label{firstpage}
\pagerange{\pageref{firstpage}--\pageref{lastpage}}
\maketitle

\begin{abstract}
In the solar wind electron velocity distributions reveal two counter-moving populations 
which may induce electromagnetic (EM) beaming instabilities 
known as heat flux instabilities. Depending on plasma parameters two distinct branches of 
whistler and firehose instabilities can be excited. These instabilities are invoked in many 
scenarios, but their interplay is still poorly understood. An exact numerical analysis is 
performed to resolve the linear Vlasov-Maxwell dispersion and characterize these two 
instabilities, e.g., growth rates, wave frequencies and thresholds, enabling 
to identify their dominance for conditions typically experienced in space plasmas. 
Of particular interest are the 
effects of suprathermal Kappa-distributed electrons which are ubiquitous in these 
environments. The dominance of whistler or firehose instability is highly conditioned by 
the beam-core relative velocity, core plasma beta and the abundance of suprathermal electrons. 
Derived in terms of relative drift velocity the instability thresholds show an inverse 
correlation with the core plasma beta for the whistler modes, and a direct correlation 
with the core plasma beta for the firehose instability. Suprathermal electrons reduce the 
effective (beaming) anisotropy inhibiting the firehose modes while the whistler instability
is stimulated. 
\end{abstract}




\section{Introduction}

Beaming electrons, also known as strahl, escape from the electrostatic potential of the Sun and 
are observed in the solar wind moving along the magnetic field lines \citep{Feldman1975, Feldman1978, 
Pilipp1987, Lin1998, Pierrard2001}. Guided by the magnetic fields these beams may also provide 
information about the magnetic field topology in interplanetary space \citep{Pagel2005}. 
The electron strahl evolves broadening its pitch-angle width \citep{Hamond1996, Pagel2007, Anderson2012} 
and loosing intensity with heliospheric distance \citep{Maksimovic2005, Stverak2009}. In the absence 
of binary collisions, the major role can be played by the wave-particle scattering, acting to broaden 
the strahl and involving, most probably, the 
enhanced fluctuations generated by the beam-plasma instabilities \citep{Gary1975,Gary1985,Scime1994,
Pistinner1998,Gary2000,Lacombe2014,Saeed2017}. Numerical simulations have confirmed the efficiency of 
magnetosonic-whistler waves \citep{Vocks2005, Saito2007}, and the electrostatic beam-driven fluctuations 
\citep{Pavan2013}. After 5.5~AU the strahl populations is largely scattered 
\citep{Graham2017}, building up the less anisotropic halo component of electrons \citep{Gurgiolo2012}, 
although signatures of this component have been detected even at 10~AU \citep{Walsh2013}.

Closer to the Sun the strahl carries the electron heat flux and the electromagnetic (EM) instabilities 
driven by this beaming component are commonly known as heat-flux instabilities \citep{Gary1975, 
Gary2000, Saeed2017, Saeed2017b}. Measured electron heat fluxes \citep{Scime1994, Bale2013} 
are lower than Spitzer-H\"arm predictions \citep{Spitzer1953}, suggesting the existence of a regulation 
mechanism for the heat flux by the kinetic beam-plasma instabilities. 
The present paper presents new results attempting to clarify the 
interplay of heat-flux instabilities, which may destabilize two distinct EM modes, namely, the 
right-handed (RH) polarized whistler (or electron-cyclotron) modes destabilized for relatively low 
beaming velocities, and the left-handed (LH) polarized electron firehose instability driven 
by more energetic beams \citep{Gary1985}. Both these two instabilities exhibit maximum growth 
rates in the magnetic field direction \citep{Gary1993}, and are usually studied in the 
proton rest frame, where the velocity distribution of electrons combines a sunward moving bulk 
or core component and a anti-sunward drifting strahl (which fulfill the zero net current condition).
Under a broad range of parameters the whistler heat-flux (WHF) instability 
has the lower beam speed threshold, and maximum growth rates higher than the firehose heat flux 
(FHF) modes \cite{Gary1985}.

Whistler beaming instability is indeed more often invoked as a plausible mechanism of  
regulation of the electron heat flux in space plasmas \citep{Gary1993, Levinson1992, Pistinner1998}. 
This is somehow surprising, given the fact that whistlers are known to be destabilized 
by an excess of kinetic (free) energy in direction perpendicular to the magnetic field (e.g., 
temperature anisotropy $T_\perp > T_\parallel$ of electrons), while parallel beams (or relative 
drifting) of electrons would be more susceptible to the nonresonant firehose-like instabilities. 
But this may explain a series of recent interpretations that question the existence of whistler 
modes \citep{Saeed2017}, and motivated also us to revisit the heat-flux instabilities. Our present investigation is intended to describe the interplay of WHF and FHF instabilities,
in particular, their regimes of dominance, and the regime of transition where both instabilities 
may co-exist and compete. The nature of instability is dictated by the velocity distributions 
of plasma particles. Thus, FHF instability is driven by the counter-beaming electrons, but 
only one of these populations is involved in the resonant excitation of whistlers. Protons are 
in general nonresonant, and WHF instability is therefore independent of the electron(-core)-proton 
temperature ratio $T_c/T_i$. However, protons may become resonant and whistlers dependent of $T_c/T_i$ 
for sufficiently high values of the electron (core) plasma beta, i.e., $\beta_{c} \equiv 
8\pi n k_B T_c/B_0^2 \gtrsim 5$ \citep{Gary2000}, a condition only marginally satisfied in the 
solar wind \citep{Stverak2008}. 

On the other hand, standard Maxwellian representation from the early studies \citep{Gary1975, Gary1985, 
Gary1993} need to be updated in accord with the observations. The electron heat flux is transported 
away from the solar corona by the suprathermal electrons with energies $E\sim 80$ eV \citep{Pagel2005} 
and well-described by a drifting-Kappa \citep{Maksimovic2005, Nieves2008, Stverak2009}. Moreover, in 
kinetic simulations the heat flux is enhanced in the presence of suprathermal beam as the power-index 
$\kappa$ decreases \citep{Landi2001}. Despite these expectations, recent studies modeling the electron 
strahl with a drifting-Kappa have found that growth rates and wave-frequencies of heat-flux 
instabilities do not vary with the power-index $\kappa$ \citep{Saeed2017}. We have revisited these
effects from a different point of view, which enabled a realistic interpretation of suprathermal 
electrons and their implication. Quantified by the lower values of $\kappa$, the abundance of suprathermals 
and their effects can be described only by contrasting with Maxwellian limit ($\kappa \to \infty$) of 
lower temperature \citep{Lazar2015, Lazar2016}. It becomes thus possible to show that 
kinetic anisotropy instabilities are stimulated by the suprathermals, confirming the additional
free energy of these populations \citep{Lazar2015, Shaaban2016, Lazar2017a, Shaaban2018}.  

The manuscript is organized as follows: In Section 2 we introduce the velocity distributions for
electrons, with two counter-moving populations reproducing the core and suprathermal strahl observed 
in the solar wind. In order to facilitate the analysis and unveil basic 
properties of heat-flux instabilities, both electron populations are considered with 
isotropic temperatures, and the effects of protons are minimized by assuming them an isotropic 
neutralizing background. The strahl is generically described by a drifting-Kappa, enabling 
to retract and complete previous results, inclusive for a standard drifting-Maxwellian strahl (in 
the limit of large power-index $\kappa \to \infty$). 
Heat-flux instabilities are discussed in detail in Sections 3 and 4, analyzing the effects of the 
beaming velocity, core beta, and the beam suprathermal population on the unstable solutions of 
whistler and electron firehose modes. In section 5 we derive the threshold conditions 
for both instabilities and identify the regimes of their dominance. 
The results are summarized in Section~4.

\section{Theory}

In the solar wind, the electron velocity distribution functions (VDFs) reveal the existence of two
components, namely, a thermal dense core and a suprathermal component drifting along the magnetic field
lines, under the influence of the field-aligned \textit{strahl} \citep{Pierrard2001, Maksimovic2005}. 
Here we adopt the VDF models in \cite{Saeed2017}, see also references therein, with a dual structure 
for the electrons combining a core (subscript c) and a beaming (subscript b) component:
\begin{equation} \label{e1}
f_e\left(v_\parallel,v_\perp \right)=\eta~f_c\left(v_\parallel,v_\perp \right)+ \delta
~f_b\left(v_\parallel,v_\perp \right), 
\end{equation} 
where $\parallel$ and $\perp$ denote (gyrotropic) directions parallel and perpendicular 
to the magnetic field, $\delta=n_b/n_0$ and $\eta=1-\delta$ are the beam-core density contrasts, respectively, 
and $n_0$ is the total electron number density. 
In a working frame fixed to protons (solar wind referential), these two populations 
are counter-moving with opposite drifting velocities, and the distribution can be interpreted as 
a superposition of a drifting-Maxwellian core with a drifting velocity $U_c$, plus a drifting-Kappa 
beam with an opposite drifting velocity $U_b$. 

Thus, for the core population we assume a drifting-Maxwellian distribution function
\begin{equation} \label{e2}
f_{c}\left( v_{\parallel },v_{\perp }\right) =\frac{1}{\pi
^{3/2}\alpha_{\perp }^{2} \alpha_{\parallel }}\exp \left(
-\frac{(v_{\parallel }+U_c)^{2}} {\alpha_{\parallel }^{2}}-\frac{v_{\perp
}^{2}}{\alpha_{\perp }^{2}}\right),   
\end{equation}
with thermal velocities $\alpha_{\parallel, \perp}$ defined by the temperature
components, as the second order moments of the distribution
\begin{subequations} \label{e3}
\begin{align}
&T_{\parallel}=\frac{m_e}{k_B}\int d\textbf{v} v_\parallel^2
f_c(v_\parallel, v_\perp)=\frac{m_e ~\alpha_{\parallel}^2}{2 k_B},\\
&T_{\perp}=\frac{m_e}{2 k_B}\int d\textbf{v} v_\perp^2
f_c(v_\parallel, v_\perp)=\frac{m_e ~\alpha_{\perp}^2}{2 k_B}.
 \end{align}
 \end{subequations} 
The beam component is described by a drifting-Kappa
\begin{align} 
f_{b}\left( v_{\parallel },v_{\perp }\right)= &\frac{1}{\pi ^{3/2}\theta _{\perp }^{2}\theta_{\parallel}}
\frac{\Gamma\left( \kappa +1\right)}{\Gamma \left( \kappa -1/2\right)} \nonumber \\
& \times \left[ 1+\frac{(v_{\parallel }-U_b)^{2}}{\kappa \theta _{ \parallel }^{2}}
+\frac{v_{\perp }^{2}}{\kappa \theta _{ \perp }^{2}}\right] ^{-\kappa-1} \label{e4}
\end{align}
with parameters $\theta_{e,\parallel, \perp}$ defined by the kinetic temperatures
\begin{align} 
T_{\parallel}^K=\frac{2 \kappa}{2 \kappa-3}\frac{m_e }{2 k_B}\theta_{\parallel}^2 
=\dfrac{2\kappa}{2\kappa-3} T_{\parallel}^M> T_{\parallel}^M, \label{e5a}\\ 
T_{\perp}^K=\frac{2 \kappa}{2 \kappa-3}\frac{me}{2 k_B}\theta_{\perp}^2 =
\dfrac{2\kappa}{2\kappa-3} T_{\perp}^M> T_{\perp}^M, \label{e5b}
\end{align}
which are assumed $\kappa-$dependent and implicitly higher than their Maxwellian limits 
\citep{Lazar2015, Lazar2017}.
Protons (subscript p) are heavier and can be assumed Maxwellian and isotropic, enabling to 
isolate the effects of electrons. The electron-proton plasma is quasi-neutral, $n_p\approx 
n_e=n_c+n_b$, with zero net current $n_c U_c+n_b U_b=0$. 

We consider the general dispersion relation of the electromagnetic (EM) modes 
propagating parallel (${\bf k} \times {\bf B}_0 = 0$) to the uniform magnetic field (${\bf B}_0$), 
e.g., in \cite{Lazar2018},
\begin{align} 
{k^2c^2 \over \omega^2} = & 1  + {4 \pi^2 \over \omega^2} 
\sum_{a=c,b,p} {e_a \over m_a} \int_{-\infty}^{\infty} {dv_{\parallel} \over \omega - k v_{\parallel} 
\pm \Omega_a} \int_0^{\infty} \, dv_{\perp} v_{\perp}^2 \nonumber \\ 
& \times \left[(\omega - k v_{\parallel}) {\partial f_{a} \over \partial v_{\perp}} + 
k v_{\perp} {\partial f_{a} \over \partial v_{\parallel}} \right]. \nonumber
\end{align}
For our three component plasma the equation reduces to
\begin{align} 
\tilde{k}^2 = &\frac{\tilde{w}}{\tilde{k} \sqrt{\mu \beta_{p}}} 
Z\left(\frac{\mu \tilde{w} \pm 1}{\tilde{k}\sqrt{\mu \beta_{p}}}\right)+
\eta~\frac{\left(\tilde{w}+u_c \tilde{k}\right)}{\tilde{k} \sqrt{\beta_c}} 
Z\left(\frac{\tilde{w} \mp 1+u_c \tilde{k}} {\tilde{k}\sqrt{\beta_{c}}}\right) \nonumber \\
&+\delta ~\frac{ \left(\tilde{w}-u_b \tilde{k}\right)} {\tilde{k} \sqrt{\beta_b}} 
Z_{\kappa}\left(\frac{\tilde{w} \mp 1-u_b \tilde{k}}{\tilde{k}\sqrt{\beta_{b}}}\right), \label{e7}
\end{align} 
where $\tilde{k}=kc/\omega_{p,e}$ is the normalized wave number, $\tilde{w}=\omega/|\Omega_e|$ 
is the normalized wave frequency, $\mu$ is the proton--electron mass contrast, $\beta_c$, 
$\beta_b$ are the core and beam plasma beta, respectively, $\delta=n_b/n_0$, 
$\eta=1-\delta$ are the beam and the core density contrast, respectively, 
$u_b=U_b~ \omega_{p,e}/(c ~ \Omega_e)$ and $u_c=\delta~ u_b/(1-\delta)$ are 
the beam and the core relative velocities,  $\pm$ denote the right-handed 
(RH) and left-handed (LH) circular polarizations, respectively, 
$Z\left(\xi^{\pm}_a\right)$ is the plasma dispersion function \citep{Fried1961}
\begin{equation}  \label{e8}
Z\left( \xi _{a}^{\pm }\right) =\frac{1}{\sqrt{\pi}}\int_{-\infty
}^{\infty }\frac{\exp \left( -x^{2}\right) }{x-\xi _{a}^{\pm }}dt,\
\ \Im \left( \xi _{a}^{\pm }\right) >0, 
\end{equation}
and $Z_{\kappa}\left(\xi^{\pm}\right)$ is the generalized modified dispersion function \citep{Lazar2008}
\begin{equation} \label{e9}
     \begin{aligned}
     Z_\kappa\left( \xi_{e}^{\pm }\right) =&\frac{1}{\pi ^{1/2}\kappa_{e}^{1/2}}\frac{\Gamma \left( \kappa_{e} \right) }{\Gamma \left(\kappa_{e} -1/2\right) }\\
     &\times\int_{-\infty }^{\infty }\frac{\left(1+x^{2}/\kappa_{e} \right) ^{-\kappa_{e}}}{x-\xi_{e}^{\pm }}dx,\ \  \Im \left(\xi _{e}^{\pm }\right) >0.
     \end{aligned}
\end{equation}
%

%
\begin{table}
	\centering
	\caption{Plasma parameters in the present study}
   \label{t1}
	\begin{tabular}{lccc} 
		\hline
		 & Beam electrons ($b$)  & Core electrons ($c$) &  Protons ($i$)\\
		\hline
		$n_j/n_i$  & 0.05 & 0.95 & 1.0\\
		$\beta_{j,\parallel}/\beta_{i,\parallel}$ & 10.0 & 1.0 & 1.0\\
		$m_j/m_i$ & 1/1836 & 1/1836 & 1.0 \\
		$\kappa$ & 3, $\infty$ & $\infty$ &  $ \infty$\\
		$\beta_{j, \perp}/\beta_{j,\parallel}$ & 1.0 & 1.0 & 1.0 \\
		\hline
	\end{tabular}
\end{table}
%

In the next we analyze the heat-flux unstable solutions of the dispersion equation \eqref{e7}.
RH solutions covert into LH modes when $\omega_r$ becomes negative, and the same convention 
applies to LH solutions. The basic set of plasma parameters used in our numerical computations 
is tabulated in Table \ref{t1}, unless otherwise noted. If we assume 
that suprathermal beam population incorporates both the halo and strahl components, the 
relative number density $\delta$ does not vary much with helioscentric distance, see Figs.~4 
and 8 in \cite{Stverak2009}, and here we adopt an average value $\delta=0.05$. For the other 
key parameters we consider values typically encountered in the solar wind at 1~AU, e.g., 
$\beta_c \simeq \beta_p \geqslant 0.1$ \citep{Stverak2008}. Lower values (e.g., $\beta_c 
\simeq \beta_p = 0.04$) may also be assumed for comparison with previous studies. 
High values of $\kappa$-index, e.g., $\kappa \geqslant 4$, are more 
specific to low heliospheric distances ($\leqslant$ 1~AU), while low values, e.g., $\kappa 
< 4$, are generated with the expansion of the solar wind beyond 1~AU \citep{Stverak2009, 
Pierrard2016}. Here we show that suprathermal effects may be noticeable even for a high 
$\kappa \geqslant 6$, usually assimilated to a Maxwellian representation \citep{Saeed2017b}.

%
\section{Instability of whistler modes}
%
In this section we study the whistler heat-flux (WHF) instability driven by the core-beam counterstreaming
electrons described above, for a wide range of parameters typically encountered in space plasmas. The 
unstable solutions are derived numerically starting from the dispersion relation (\ref{e7}) for 
RH modes (for $\xi_p^+$). We consider two alternative representations for the beam component. 
First, we assume both counter-beaming populations Maxwellian-distributed, which is a common approach 
in the literature, see \cite{Gary1985, Saeed2017}, and references therein. In this case the beam is a 
drifting-Maxwellian recovered from a drifting-Kappa in Eq.~(\ref{e4}), in the limit $\kappa\rightarrow 
\infty$. In the second part we will consider a scenario more realistic for the solar wind conditions, 
when the electron beam is reproduced by a drifting-Kappa and instability conditions may be altered by 
the suprathermal electrons \citep{Lazar2016}.

%
\subsection{Maxwellian beam}
%
\begin{figure}
\centering 
\includegraphics[width=20pc]{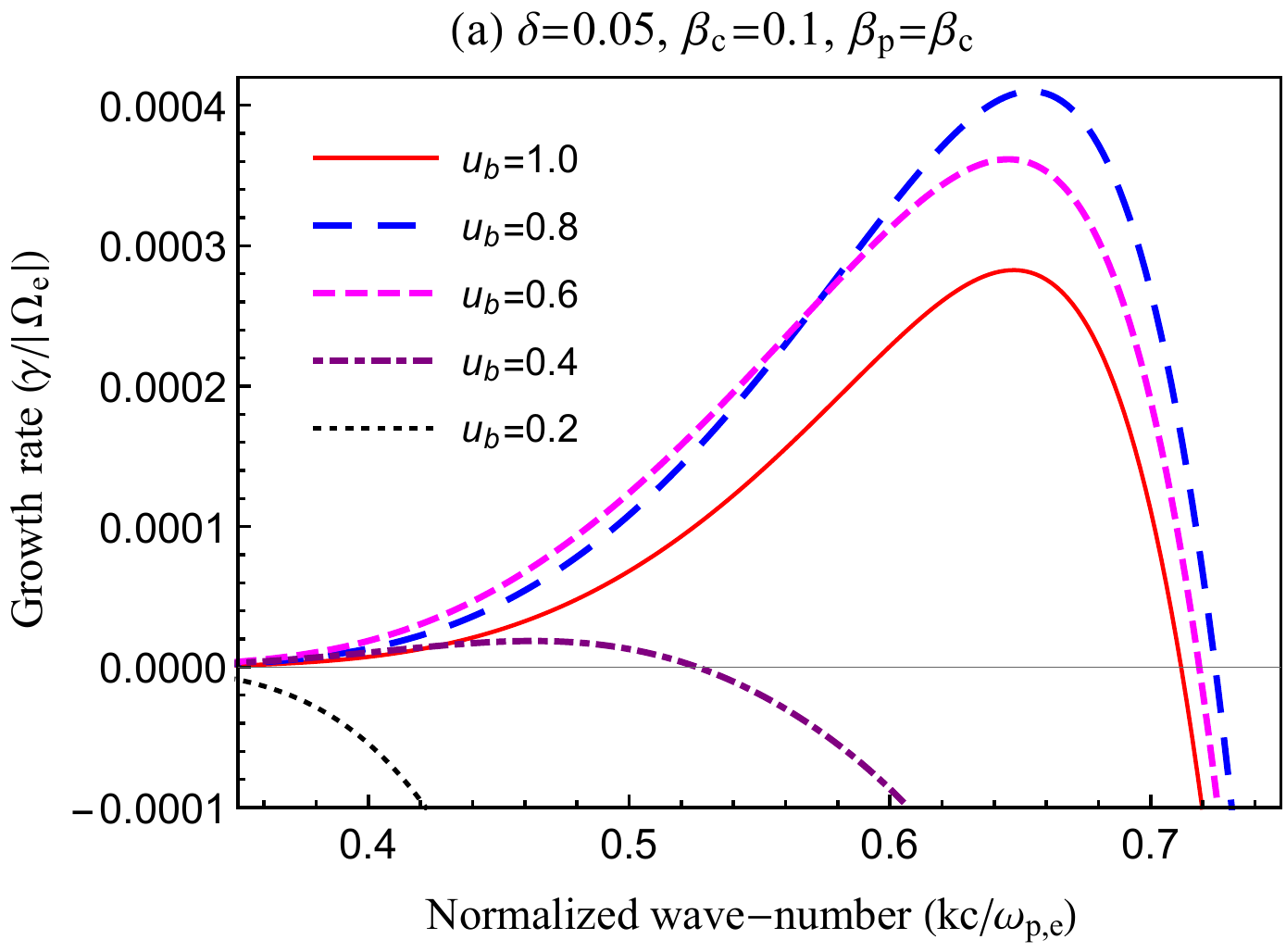}
\includegraphics[width=20pc]{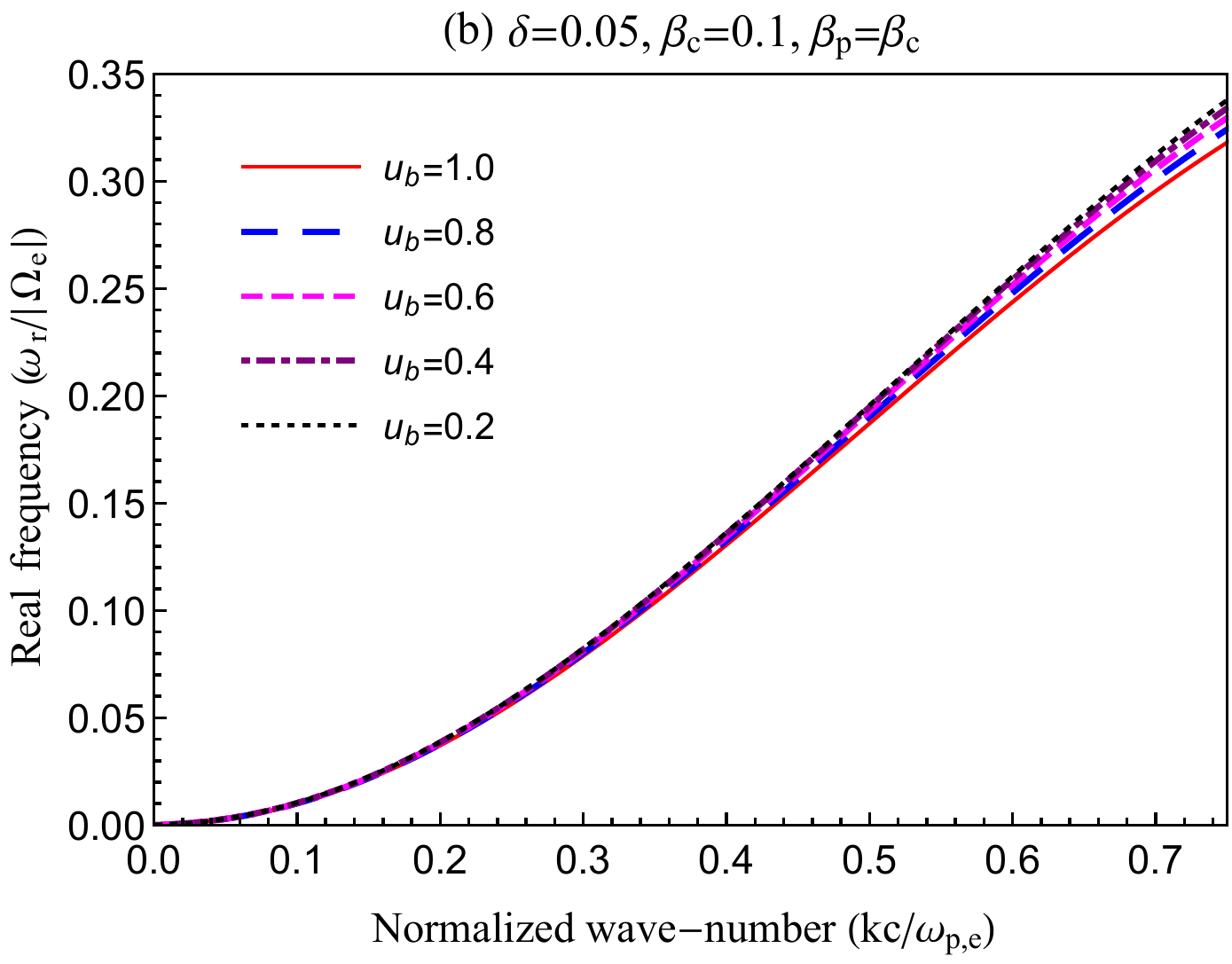}
\caption{Effect of the beam velocity $u_b=[0.2-1.0]$ on the WHF instability growth rate 
(panel a) and wave-frequency (panel b).}
\label{fig:1}
\end{figure}
%
\begin{figure}
\centering 
\includegraphics[width=20pc]{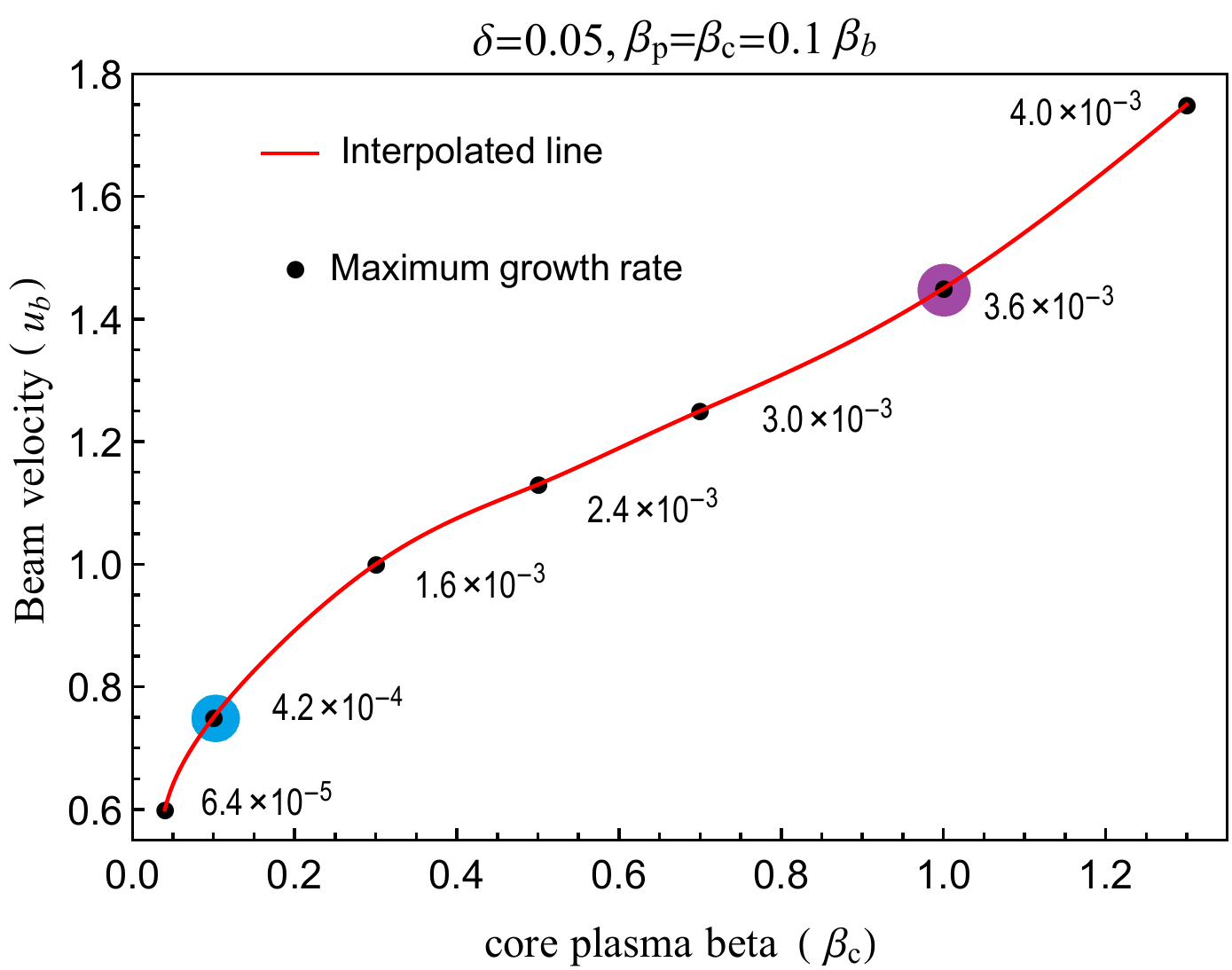}
\caption{Maximum growth-rates conditions in terms of the beam speed and the core plasma beta. }
\label{fig:2}
\end{figure}
%

\begin{figure}
\centering 
\includegraphics[width=20pc]{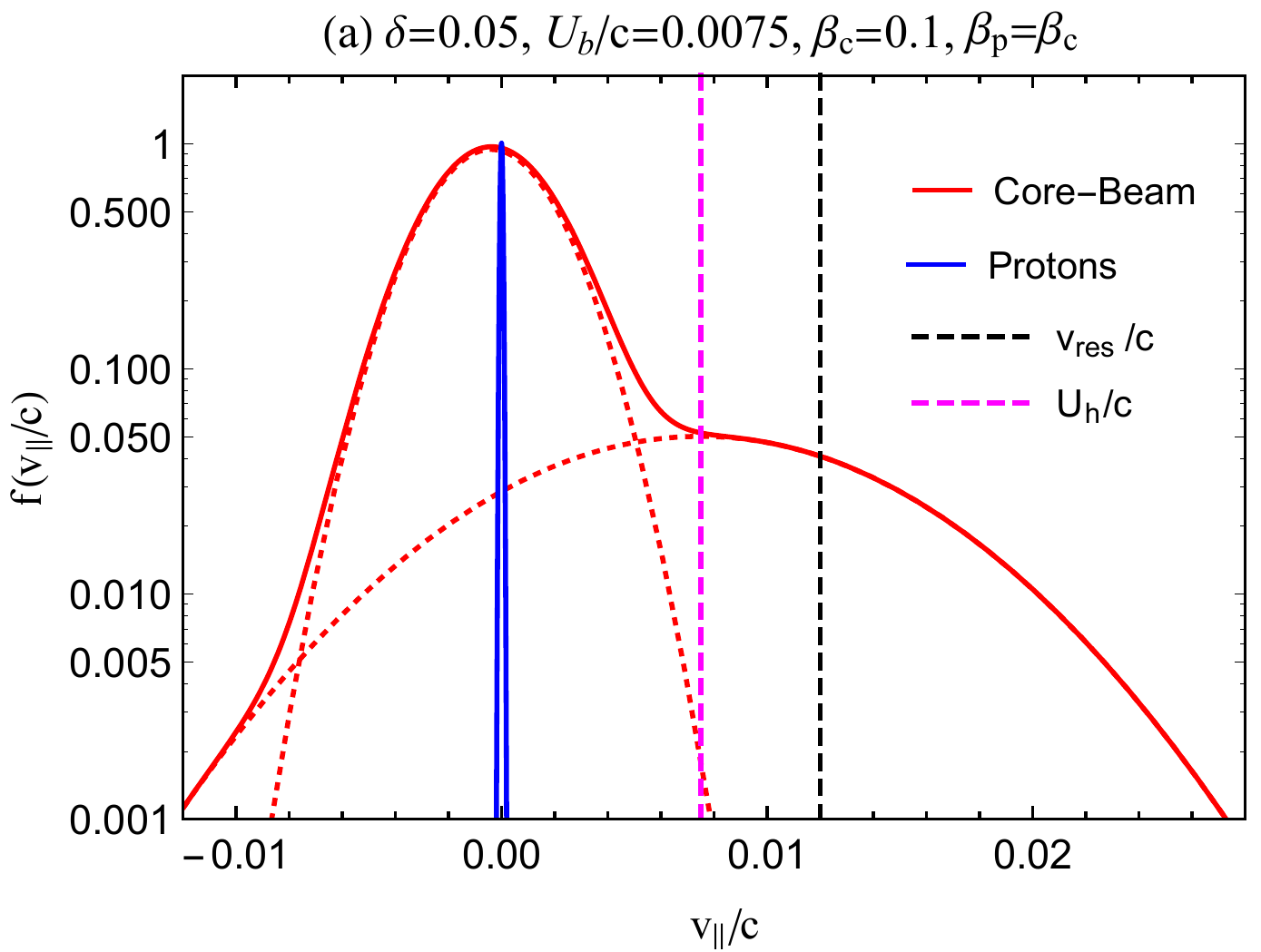}
\includegraphics[width=20pc]{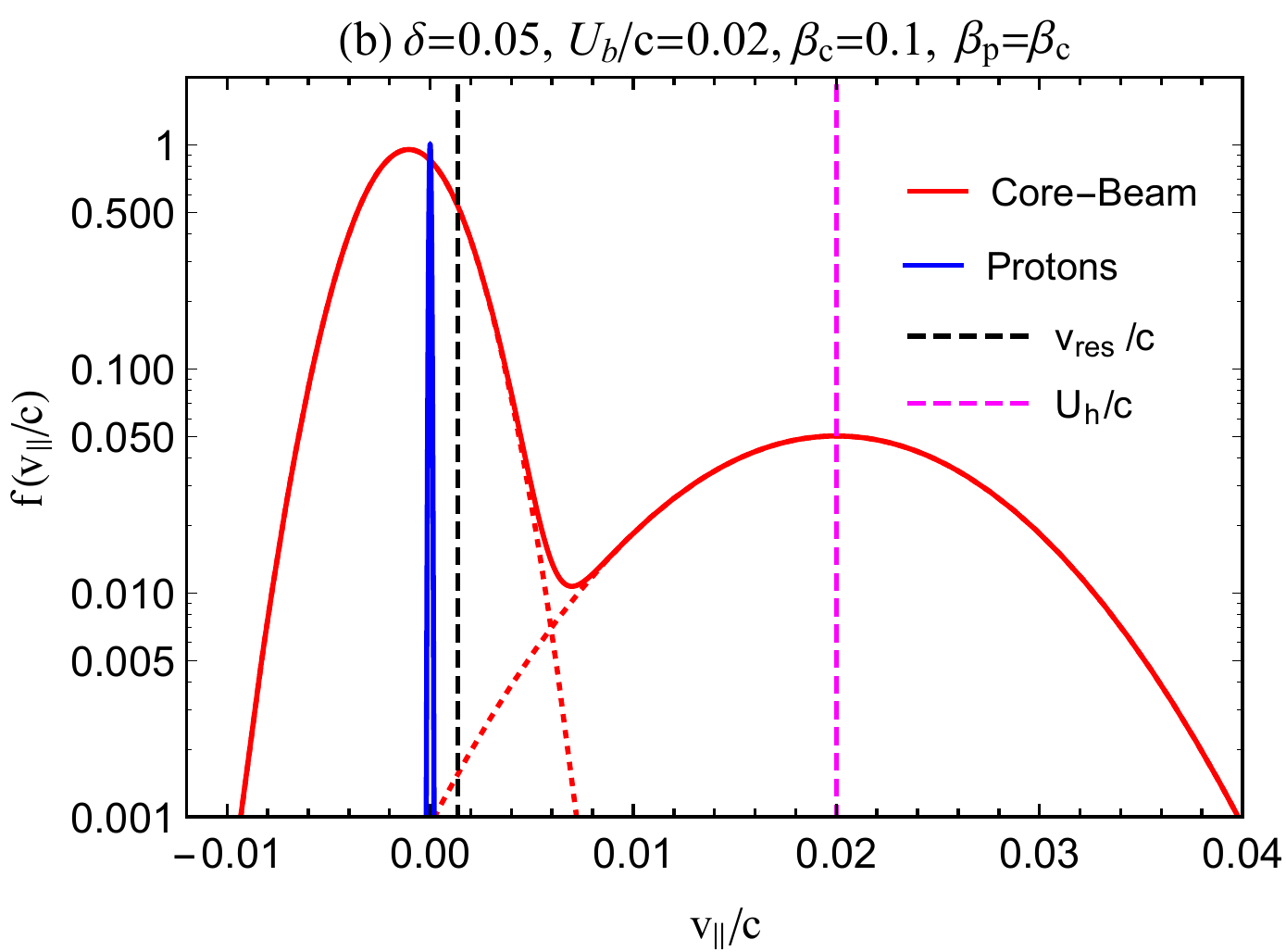}
\caption{Velocity distributions susceptible to the WHF instability for $u_b=0.75$ in panel (a), and 
$u_b=2.0$ in panel (b).}
\label{fig:3}
\end{figure}

Fig.~\ref{fig:1} describes the influence of different beaming velocities $u_b=$1.0, 0.8, 0.6, 
0.4, 0.2 on the WHF growth rates (panel a) and wave frequency (panel b), assuming $\delta=0.05$ 
and $\beta_c=\beta_p=0.1$. Growth rates show a non-uniform variation, increasing and then 
decreasing with increasing $u_b$, and maximum growth rate (highest peak) is reached for 
$u_b\simeq0.8$. The corresponding wave-frequencies 
in Fig.~\ref{fig:1}(b) are positive $\omega_r>0$, i.e., RH polarization, and slightly decrease 
with increasing the beaming velocity. In these examples, the instability is driven by the 
relative counterstreaming motion of electron components \citep{Gary1985}. In order to 
identify the dominance regime of WHF instability, in Fig.~\ref{fig:2} we show the beam 
velocity threshold $u_{bt}$ required for the instability to display maximum growth rates, 
as a function of the core plasma beta $\beta_c$. Physical conditions are provided 
for the fastest growing modes, explicitly indicating the maximum growth rates $\gamma_m/ 
|\Omega_e|$ which increase with $\beta_c$ and $u_b$, as also the electrons become 
resonant. Moreover, Fig.~\ref{fig:2} shows that the beam velocity 
threshold $u_{bt}$ increases as the core beta increases. The correlation between growth rates 
and the driver, e.g., beaming velocity $u_b$, is suggested by the resonance condition: 
whistlers are destabilized by the resonant electrons satisfying $|\xi_b^+| \simeq 1$, 
which implies $|\Omega_e| = k V_{th,res} - k U_b > 0$. Thus, the resonant instability 
requires $U_b < V_{th}$, and Fig.~\ref{fig:3} presents two counter-beaming electron 
distributions (red lines) satisfying this condition, and nonstreaming Maxwellian protons 
(blue lines). Plasma parameters are the same as in 
Fig.~\ref{fig:1}, excepting the normalized beaming velocity $u_b=~U_b ~\omega_e/(c~|\Omega_e|)
=0.75$ (required for the instability to display maximum growth rate) in panel (a) and $u_b=2.0$ 
in panel (b) ($\omega_e/|\Omega_e|=100$). For the core-beam electrons we chose 
$\alpha_\parallel/c=\sqrt{\beta_c} ~\omega_e/|\Omega_e|=0.0032$ ($\beta_c=0.1$) and 
$\theta_\parallel/c=\sqrt{\beta_b} ~\omega_e/|\Omega_e|=0.01$ (with $\beta_b=1.0$), while for 
the stationary protons $\alpha_\parallel/c=\sqrt{\beta_p/\mu} ~\omega_e/|\Omega_e|= 
7.3\times 10^{-5}$ ($\beta_p=\beta_c=0.1$). For a beaming velocity $u_b=0.75$ the resonant 
electrons have $v_{res}=V_{res}~\omega_e/(c~|\Omega_e|)=1.2$ (panel a), while for $u_b=2.0$ we 
find $v_{res}=0.136$ (panel b). The resonant velocity $v_{res}$ in panel~(a) is slightly higher 
than the beaming velocity $u_b$ involving more electrons from the beam population and enhancing 
the instability. The unstable solution in this case corresponds to the blue shaded point in 
Fig.~\ref{fig:2} with maximum growth rate $\gamma_m=$4.2$\times$10$^{-4}$ $|~\Omega_e|$. 
By contrast, in panel~(b) the resonant electrons with $v_{res}=0.136$ (much 
lower than $u_b=2.0$) are located in the core, and (maximum) growth rate is reduced to 
$\gamma_m=$2.5$\times$10$^{-6}$ $|~\Omega_e|$.

%
%
\begin{figure}
\centering 
\includegraphics[width=21pc, trim={0 0 14cm 0},clip]{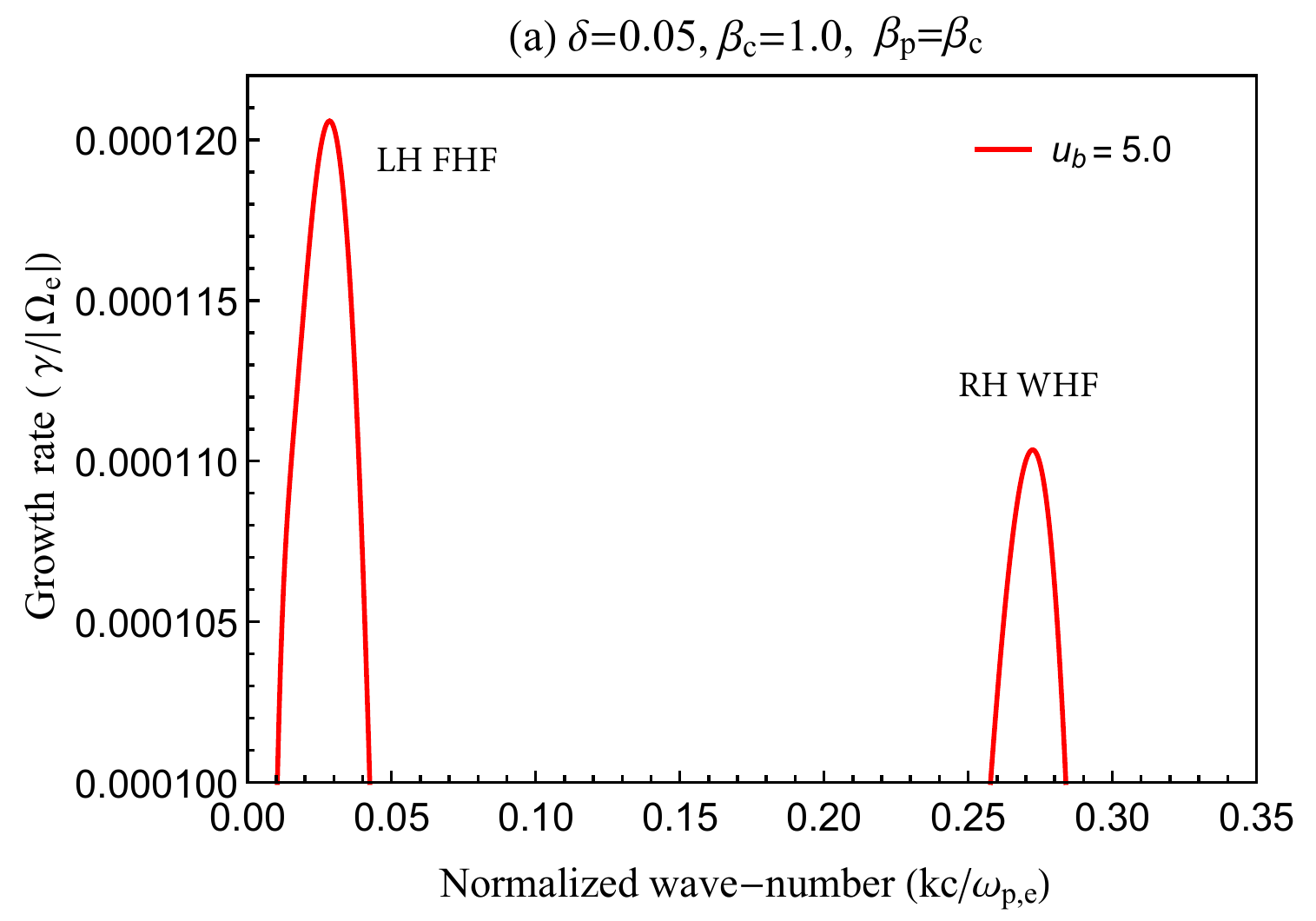}
\includegraphics[width=20pc, trim={15cm 0 0 0},clip]{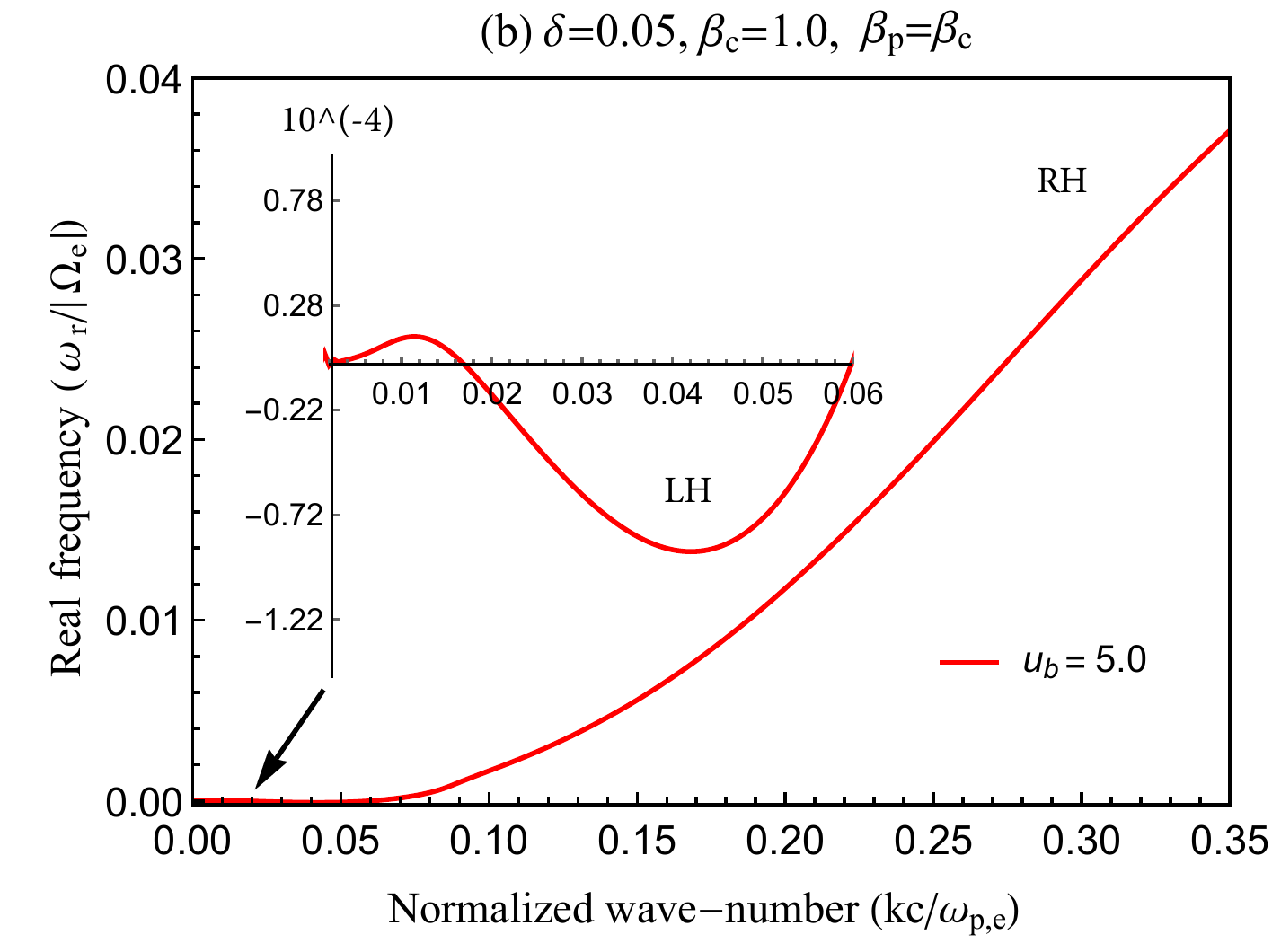}
\caption{Transitive regime with both peaks of RH WHF and LH FHF growth rates in panel (a),
and different polarizations indicated by the wave frequency in panel (b).}
\label{fig:4}
\end{figure}
\begin{figure}
\centering 
\includegraphics[width=15.5pc]{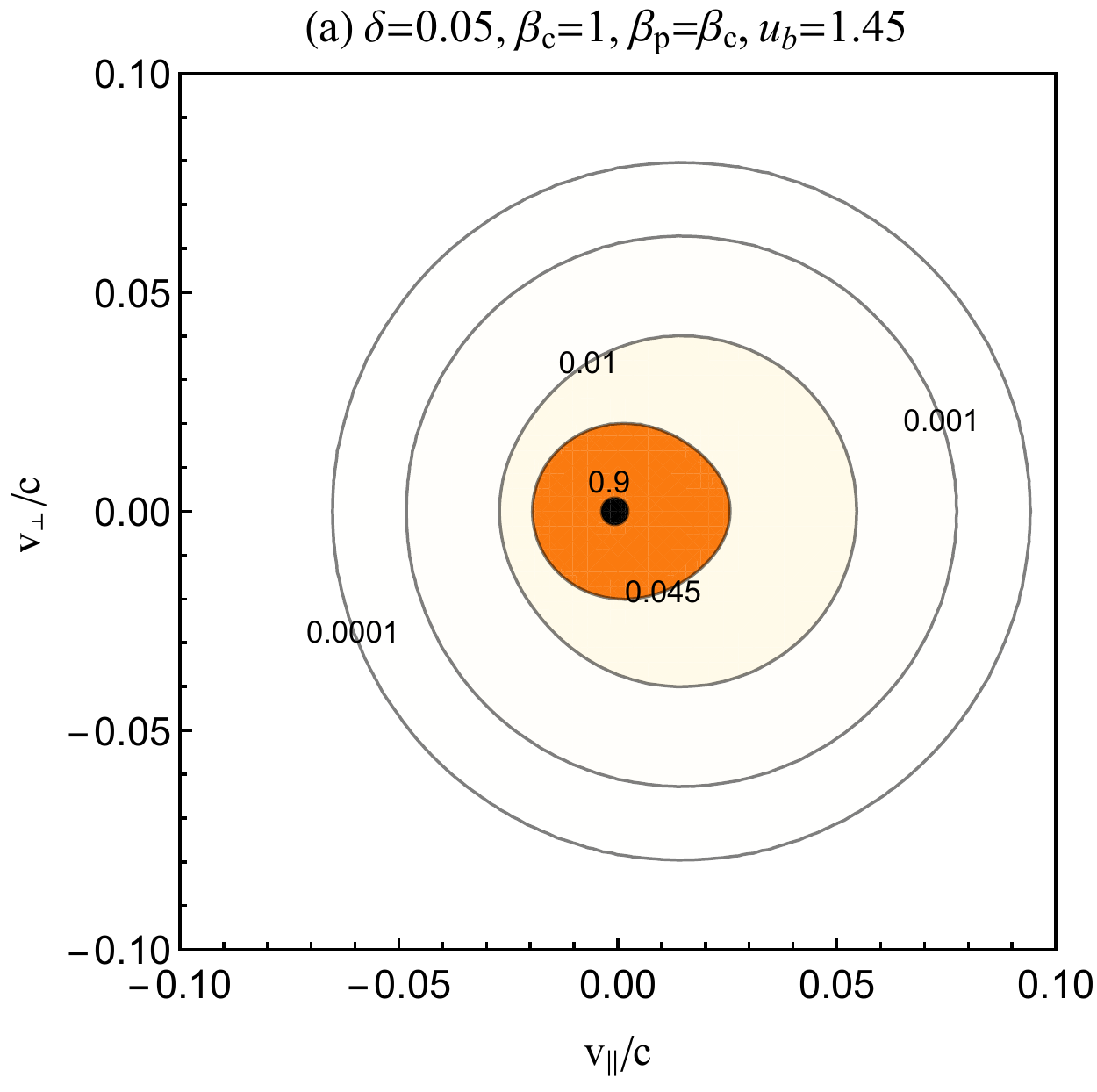}
\includegraphics[width=15.5pc]{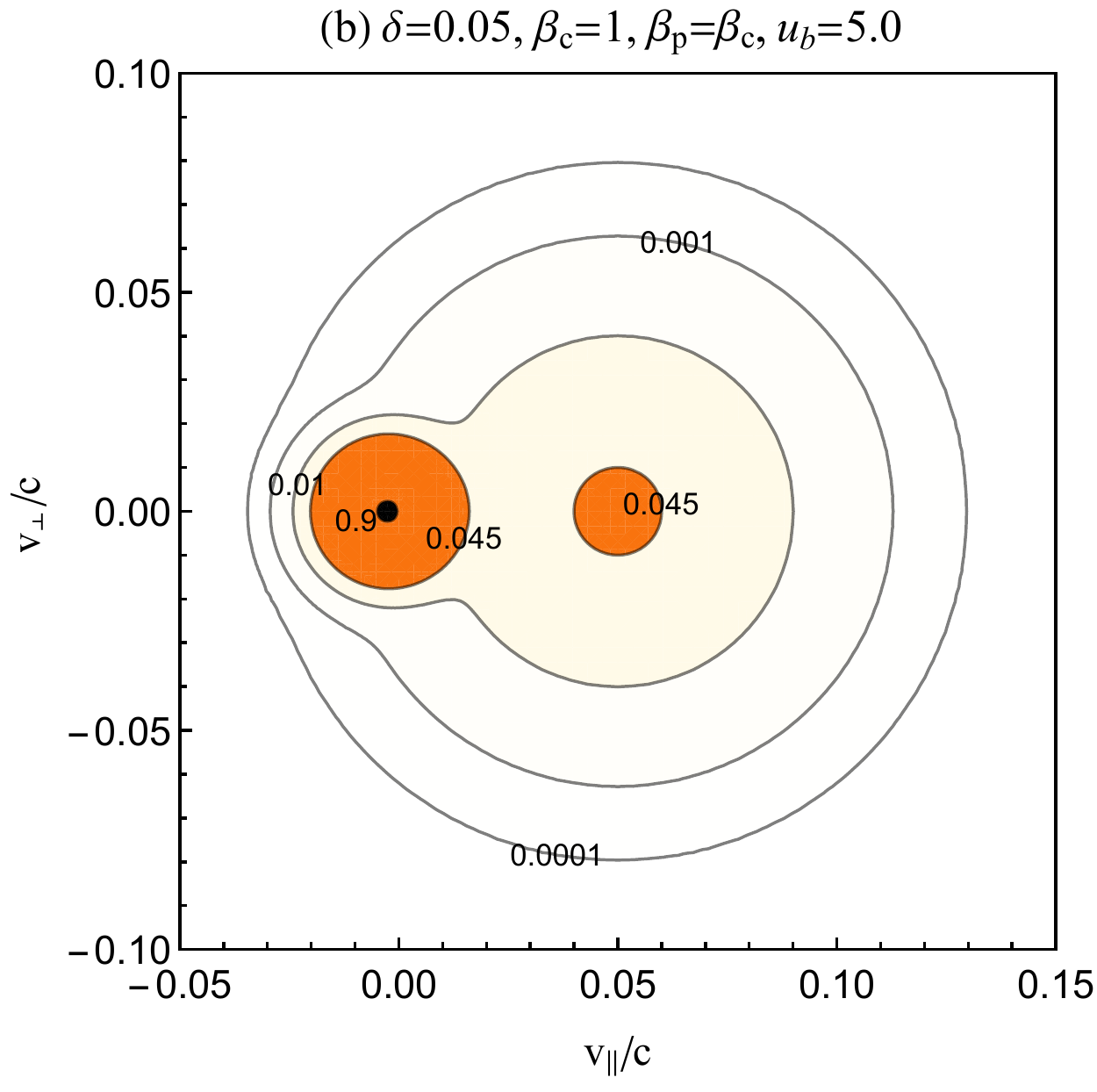}
\caption{Beam-plasma distributions susceptible to the WHF instability ($u_b = 1.5$) in panel 
(a), and the FHF instability ($u_b = 5.0$) in panel~(b).}
\label{fig:5}
\end{figure}
%

At this stage, an important question should be answered. The non-uniform variation of 
maximum growth rates (which increase and then decrease) with increasing beaming velocity 
$u_b$, prompt us to question the role played by this instability in the relaxation of 
more energetic beams with $u_b > u_{bt}$. Fig.~\ref{fig:4} presents growth rates (panel a)
and wave frequency (panel b) of HF instabilities for the same plasma parameters as in 
Fig.~\ref{fig:2}, for $\beta_c=1.0$ (purple shaded point), but for a higher beaming 
velocity $u_b=5$. In this case the growth rate displays two distinct peaks, one of the 
electron firehose heat-flux (FHF) instability at low wave numbers, and a second peak 
corresponding to WHF instability at larger wave-numbers. The wave frequency in panel (b) 
confirms the LH polarization of the FHF modes (the zoom-in subplot), which converts to a RH 
polarization of the WHF peak. If the density contrast $\delta$ is constant, this transition 
between the WHF and FHF instabilities depends only on the plasma beta $\beta_c$ and the 
beaming velocity $u_b$. Similar transition  have been reported by \citet{Gary1984} between 
the LH and RH modes driven unstable by the EM ion beam instabilities (RH nonresonant ion 
beam instability obtained for low beam-core temperature contrast $T_b/T_c=~10$, converts to 
a resonant LH mode for higher $T_b/T_c=~100$, see Figs.~2 and 7 in \citet{Gary1984}). The 
interplay between these growing modes will be discussed here below after identifying the 
FHF instability conditions in section~4. 

In order to understand the transition between the WHF and FHF instabilities, in Fig.~\ref{fig:5} 
we display two distinct VDFs, in panel (a) the one used to derive the WHF solution corresponding 
to the purple shaded point in Fig.~\ref{fig:2}, and in panel (b) the VDF at the origin of two
peaks in Fig.~\ref{fig:4}. In panel~(a) the distribution is typical for WHF solutions with 
maximized growth rates, while panel~(b) shows a distribution relevant for the transition between 
WHF and FHF, as already described in Fig.~\ref{fig:4}. These contours are plotted for different 
beaming velocities $u_b=1.45$ in panel (a) and $u_b=5.0$ in panel (b), but for the same plasma
beta parameters, i.e., $\beta_c=1.0$ and $\beta_b=10.0$. For a small $u_b=1.45$ the contrast 
between core and beam is modest, see the contour level 0.045 (orange color) in panel (a), and 
these components can be considered 'strongly coupled', as a singular population. 
For a higher $u_b=5.0$ in panel (b), the beam displays a peak markedly departed from the 
core peak, see the contour level 0.045 (orange color). In this case, the beam appears 
'weakly coupled' and any further increase of beaming velocity in parallel direction 
implies an effective increase of the anisotropy in this direction, which must be favorable to 
a LH FHF instability. Panels~(a) in Fig.~\ref{fig:3} and \ref{fig:5} show the distributions 
corresponding to the blue ($\beta_c=0.1$) and purple ($\beta_c=1.0$) shaded points in 
Fig.~\ref{fig:2}. A comparison becomes straightforward and should explain the increase of 
the beaming velocity threshold $u_{bt}$ with $\beta_c$. In Fig.~\ref{fig:5}~(a) 
the electrons have thermal velocities $v_{\parallel}/c$ much higher than those in 
Fig.~\ref{fig:3}~(a) (assuming the plasma beta increasing with thermal velocity), 
and WHF instability is excited for higher beaming velocities, i.e. $u_b=1.45$.

\subsection{Kappa distributed beam}
%
\begin{figure}
\centering 
\includegraphics[width=20pc]{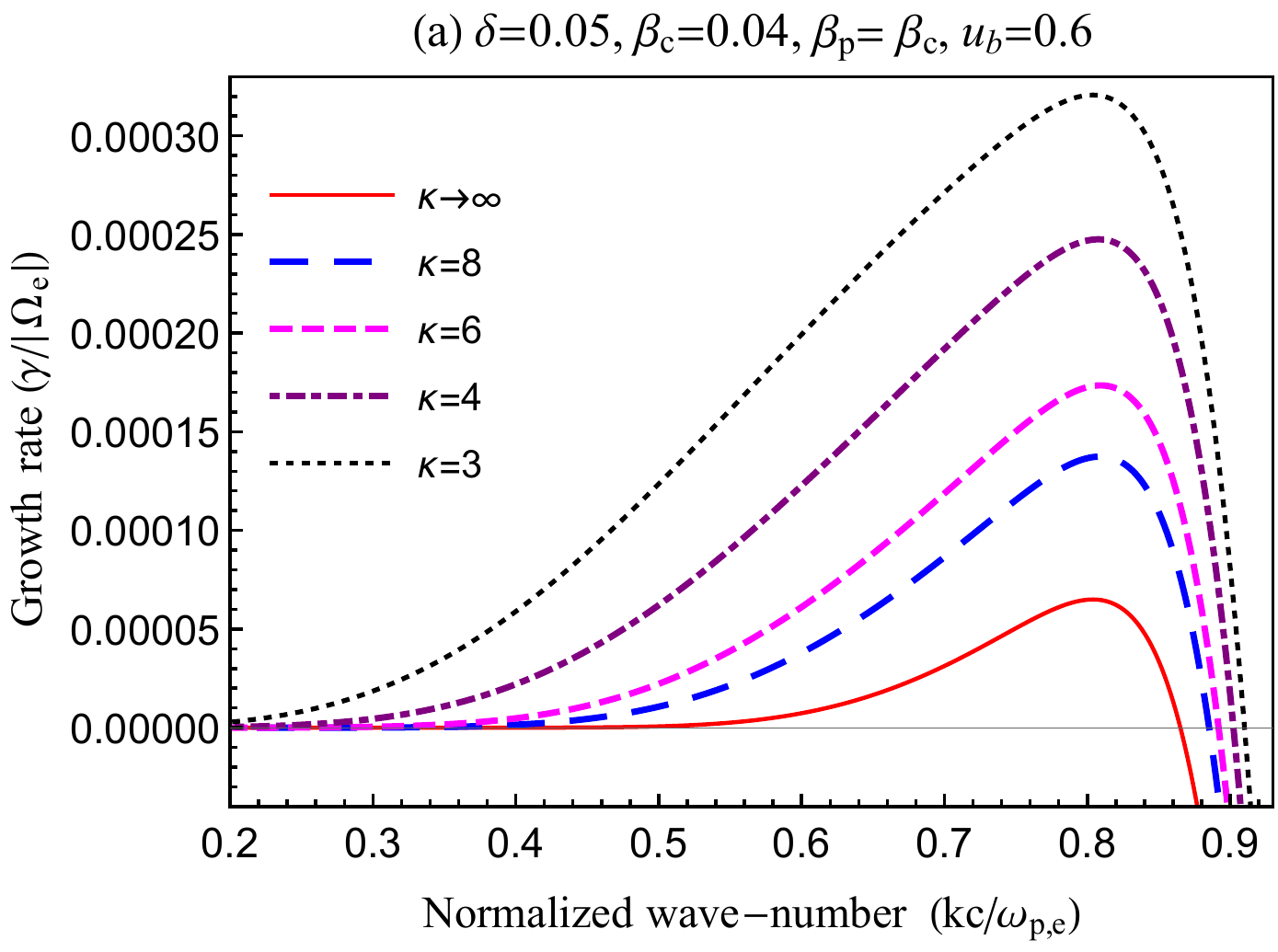}
\includegraphics[width=20pc]{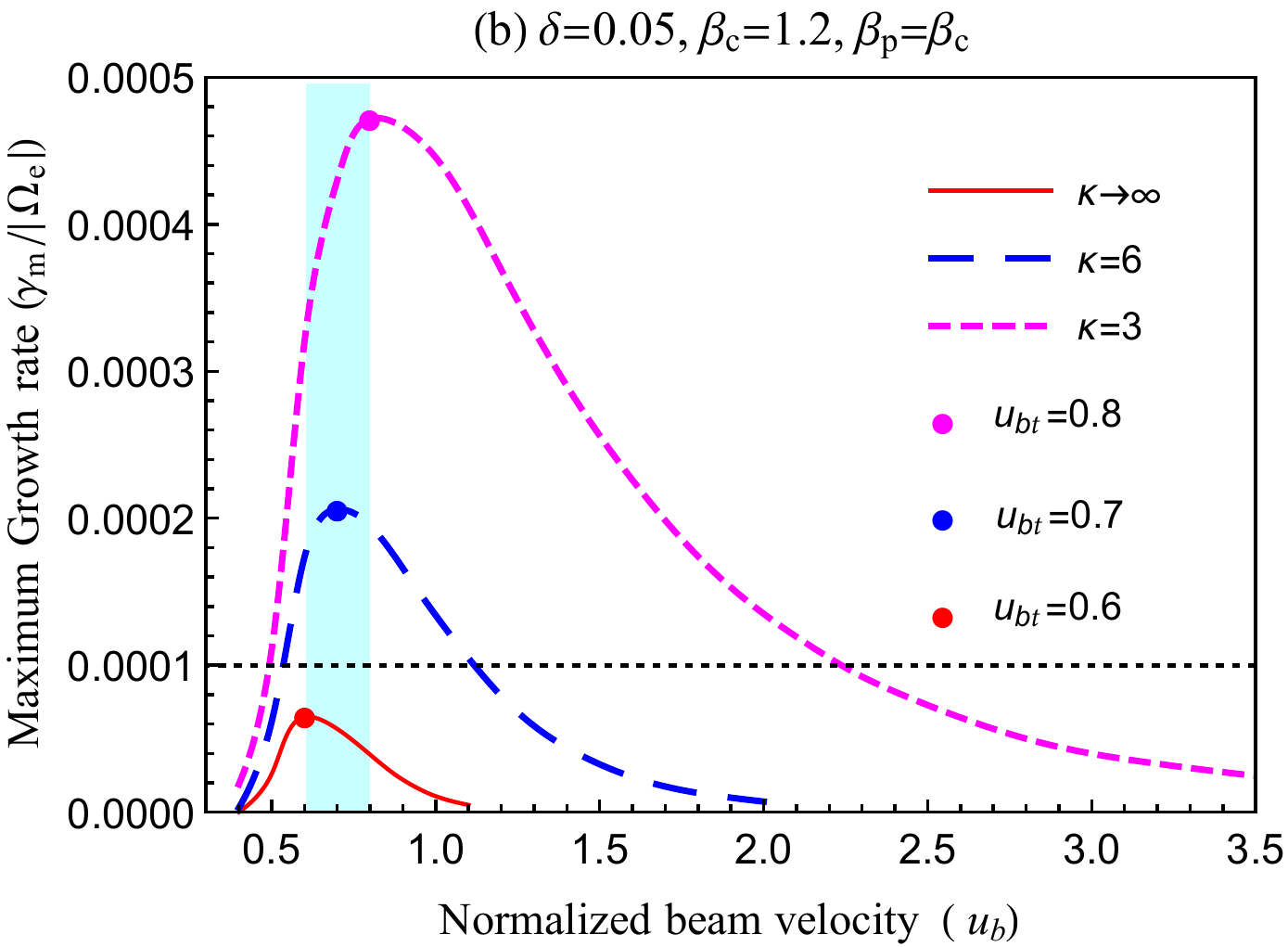}
\caption{Effect of the power-index $\kappa$ on the WHF instability: growth rate in panel (a), 
and maximum growth rates (as a function of the beam velocity $u_b$) in panel (b).}
\label{fig:6}
\end{figure}
%

In a more realistic solar wind scenario the suprathermal beam population is better 
reproduced by a drifting-Kappa, as given in Eq.~\eqref{e4}.
Fig.~\ref{fig:6}~(a) describes the effect of suprathermal electrons on the WHF solutions 
by varying the power-index $\kappa=3, 4, 6, 8, \infty$. The other parameters (kept constant 
in the analysis) are $u_b=0.6$, $\beta_c=0.04$, $\beta_p=\beta_c=\beta_b/10$, and density 
contrast $\delta=0.05$. The suprathermal electrons in the beam have 
a stimulating effect on the WHF, enhancing growth rates and increasing the range of unstable wave 
numbers. These results are in agreement with kinetic simulations, which suggest that the heat 
flux in the outer corona and solar wind is stimulated by the electron suprathermal populations 
\citep{Landi2001}. The wave-frequency (not shown here) remains roughly unchanged to the 
variation of $\kappa$. Panel~(b) provides a more general picture on the maximum growth rates
varying with the beaming velocity threshold $u_{bt}$ and the power-index $\kappa = 3, 6, \infty$ 
(for $\beta_c=0.04$ and $\delta=0.05$). The maximum growth rates are markedly enhanced with 
the abundance of suprathermal population in the beam, i.e., with decreasing $\kappa$. For 
$\kappa=3$ the maximum growth rate exhibits a peak five times greater than that obtained for 
Maxwellian ($\kappa\rightarrow\infty$) counterstreams. As a consequence of that, 
conditions favorable to WHF instabilities extend to markedly higher beaming velocities, 
roughly four times higher than that needed by Maxwellian electron beams to stabilize. 
Beaming velocity thresholds $u_{bt}$, associated to different maximum growth rates, increase 
in the presence of suprathermal beaming electrons, i.e., decreasing $\kappa$ (see the cyan 
shaded area).

\section{Beaming firehose instability}

In this section we investigate the second branch of HF instabilities, namely, the beaming firehose, 
also known as the firehose heat-flux (FHF) instability, which develops for higher beaming 
velocities $u_b$. The unstable modes are LH polarized and can be obtained from a conversion
of RH modes with increasing $u_b$ (see the analysis in Fig.~4), or starting directly from the 
dispersion relation (\ref{e7}) for LH modes ($\xi_p^-$).
 
\subsection{Maxwellian beam}
%
\begin{figure}
\centering 
\includegraphics[width=20pc]{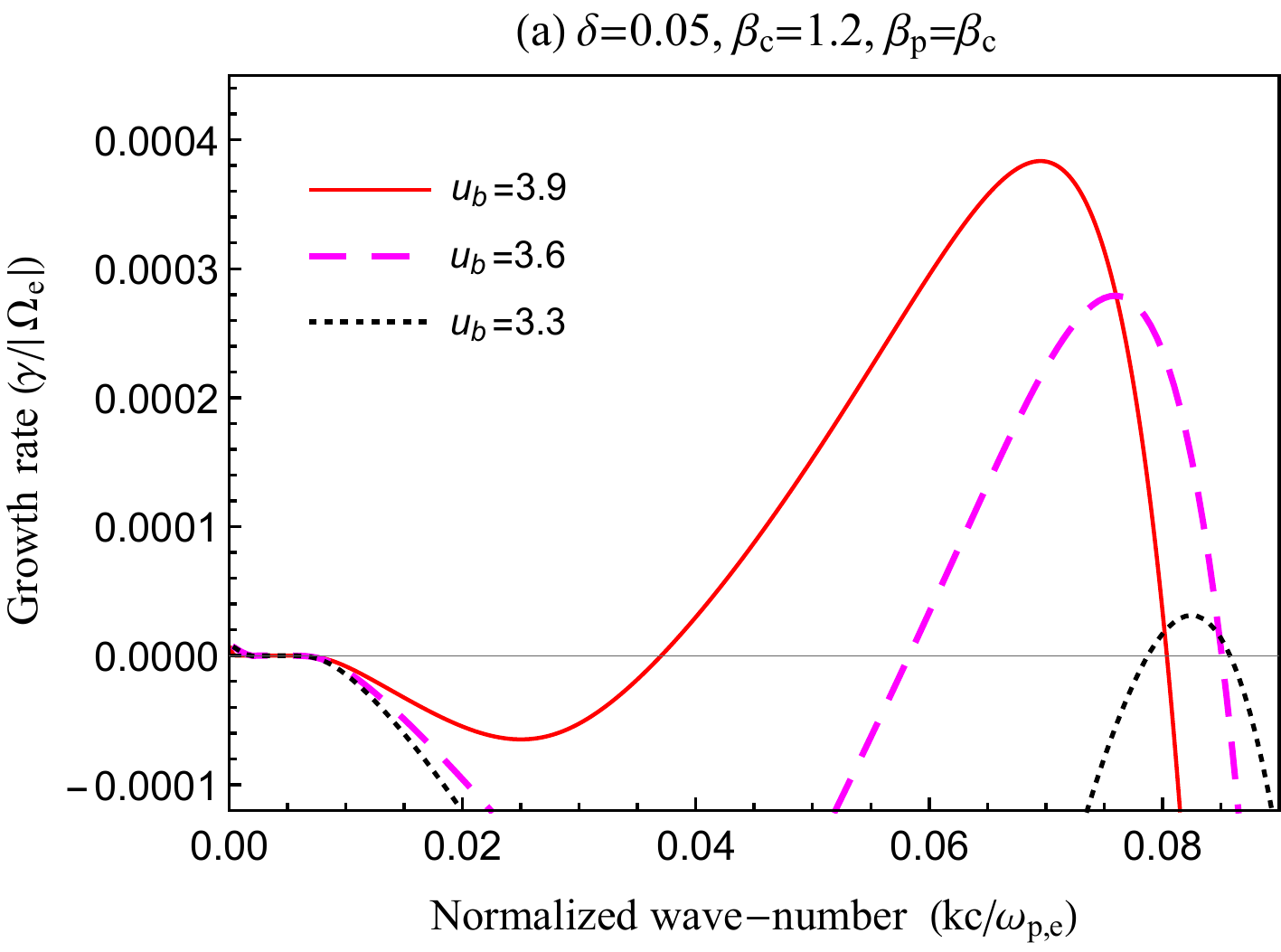}
\includegraphics[width=20pc]{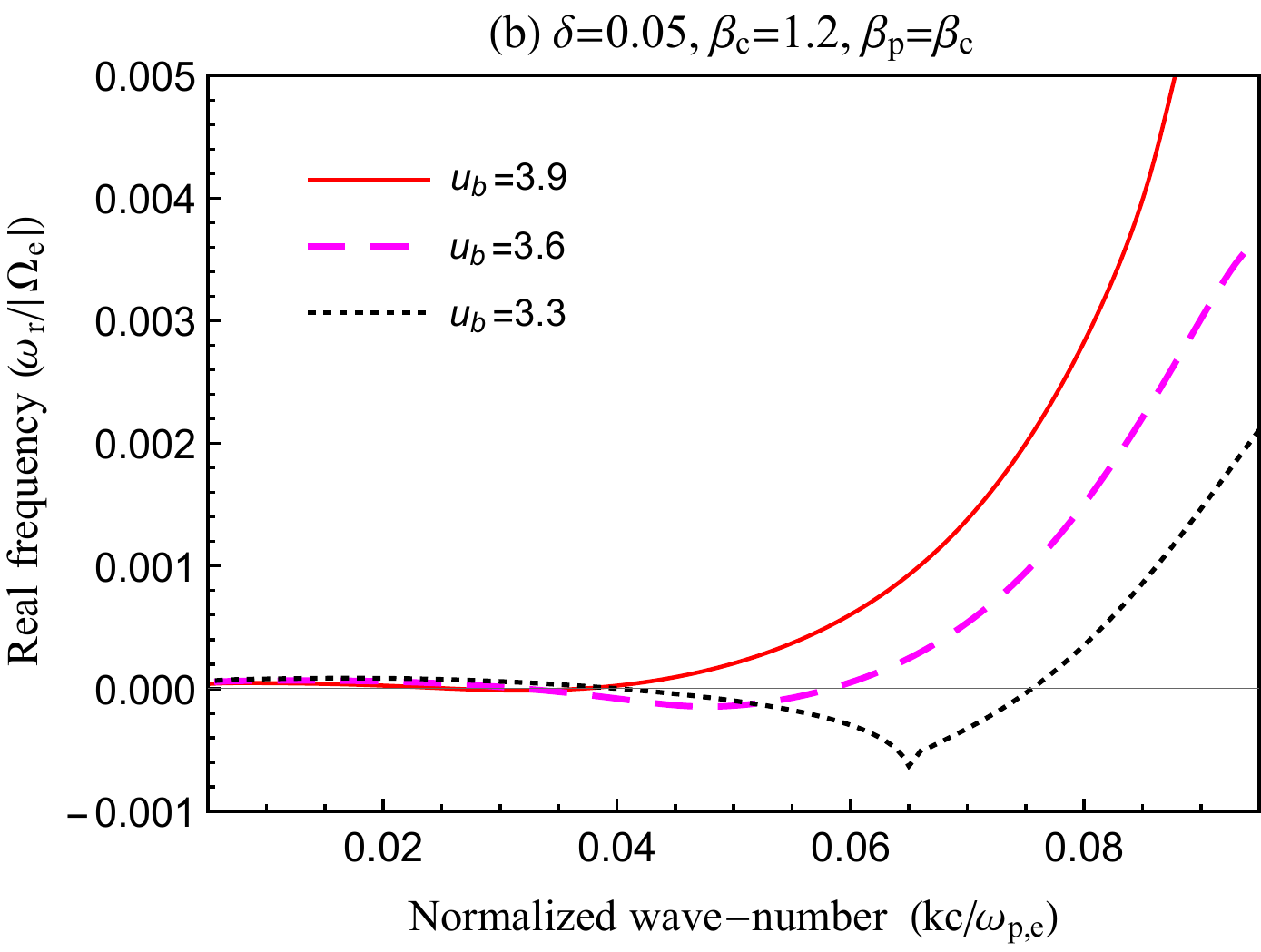}
\caption{Effect of the beaming velocity $u_b$ on the FHF instability: growth rates in panel (a) and wave 
frequency in panel (b).}
\label{fig:7}
\end{figure}

We start again with an analysis of an idealized situation when both counter-beams of electrons 
are Maxwellian. The dispersion relation (\ref{e7}) for LH modes is solved numerically in the 
limit of $\kappa \to \infty$. From the variation with beaming velocity we can identify 
the instability conditions and mode polarization (given by the sign of $\omega_r$). 
Recent studies of FHF instability have restricted to a low $\beta$ regime i.e., $\beta_c=0.04$ 
\citep{Saeed2017}, although firehose instability is significantly more efficient for
high $\beta_c>1$ regimes. In Fig.~\ref{fig:7} we present FHF solutions for a relatively 
high core plasma beta $\beta_c=1.2$. Panel~(a) shows growth rates increasing as the 
beaming velocity $u_b$ increases, while the range of unstable wave numbers increases
towards lower values. By comparison to whistlers these FHF modes are destabilized at
lower wavenumbers (i.e., higher proton scales). The corresponding wave frequencies in panel (b) are 
markedly increased by increasing $u_b$, and remain positive $\omega_r>0$ in the range 
of FHF peaks. The conversion from RH modes is still visible for the less energetic
beams, i.e., for $u_b = 3.3$.

\subsection{Kappa distributed beam}

\begin{figure}
\centering 
\includegraphics[width=20pc]{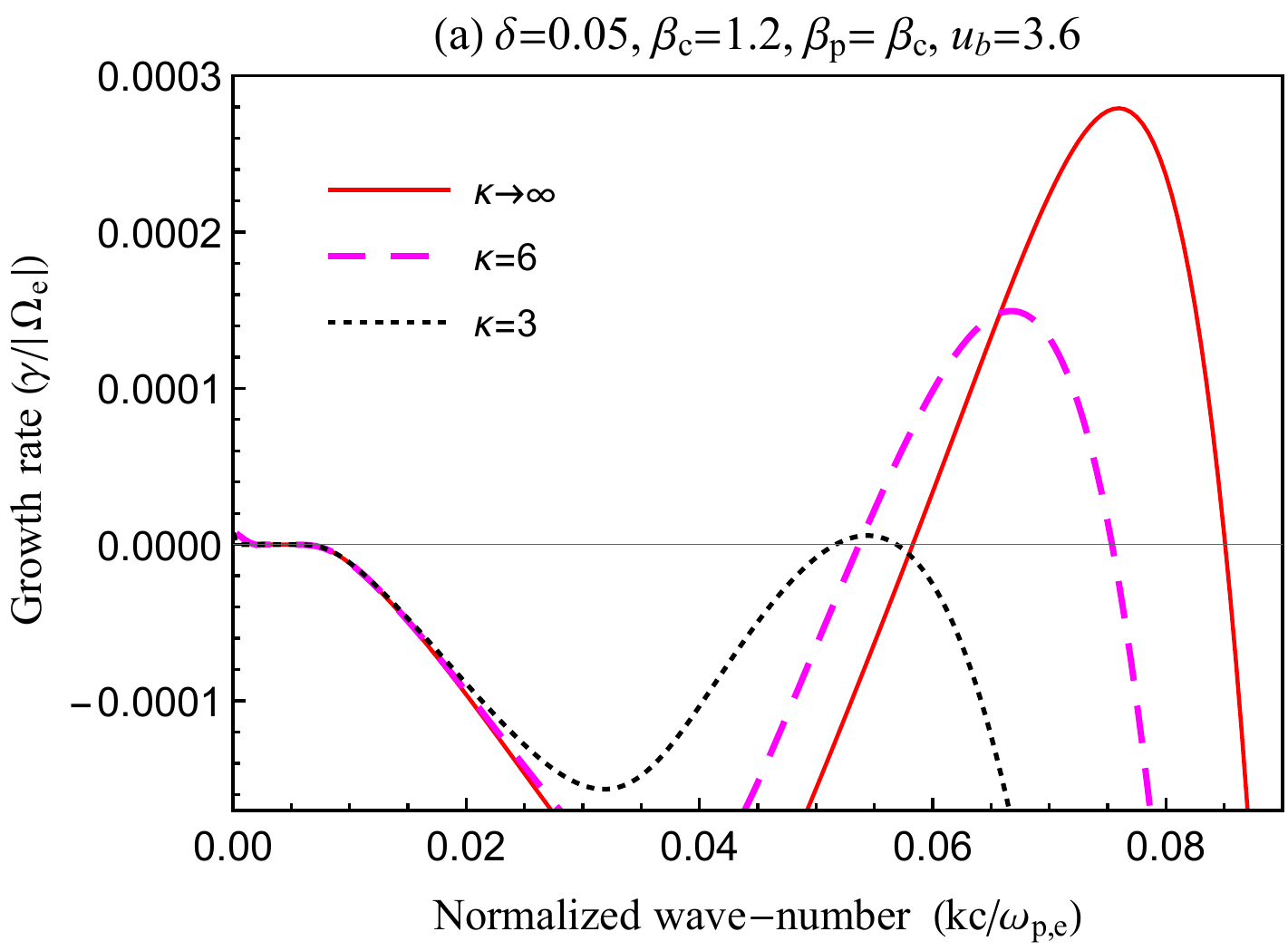}
\includegraphics[width=20pc]{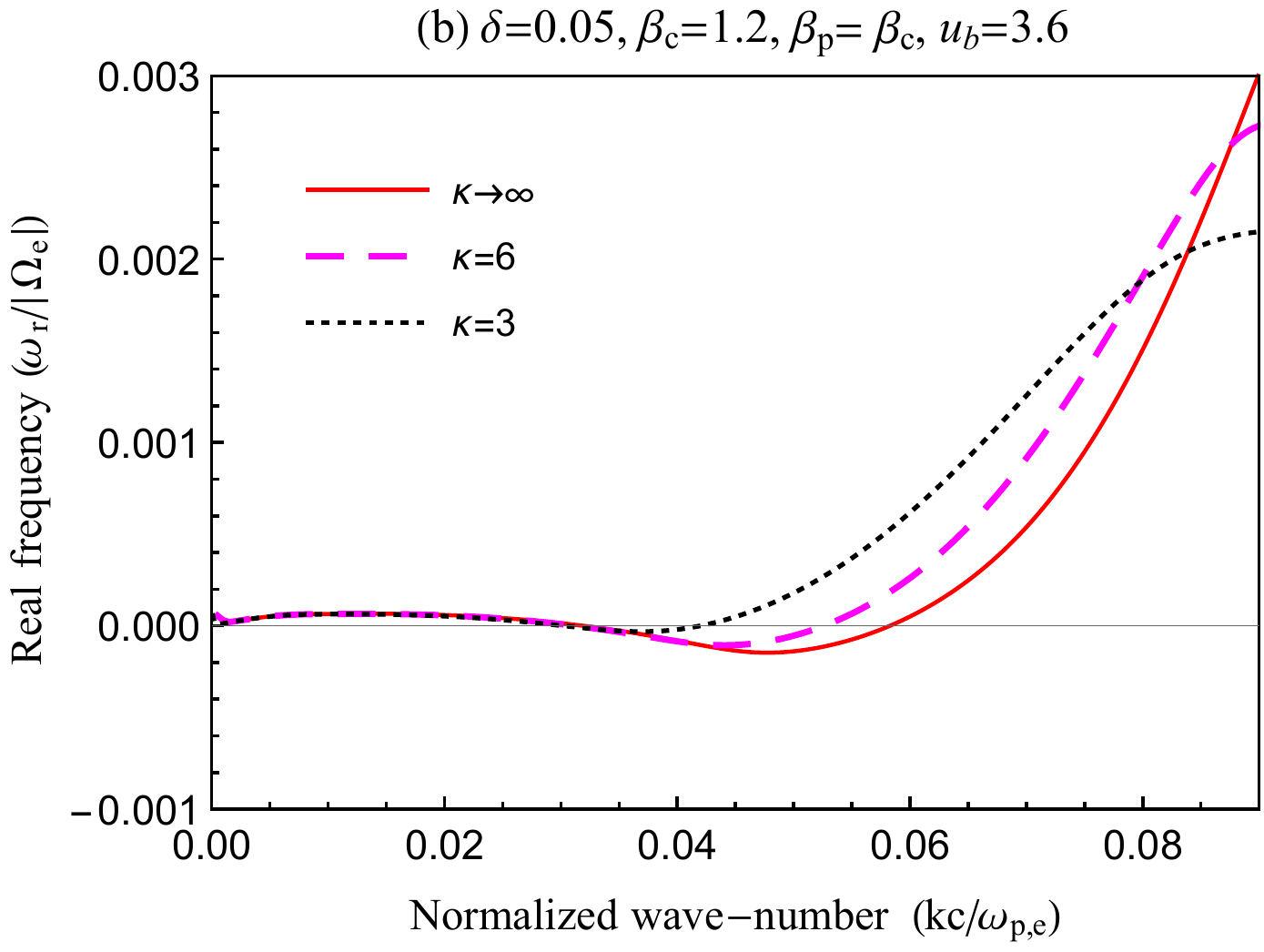}
\caption{Effect of the power-index $\kappa$ on the FHF instability: growth rate in panel (a), 
and wave frequency in panel (b).}
\label{fig:8}
\end{figure}
%
\begin{figure}
\centering 
\includegraphics[width=15.pc]{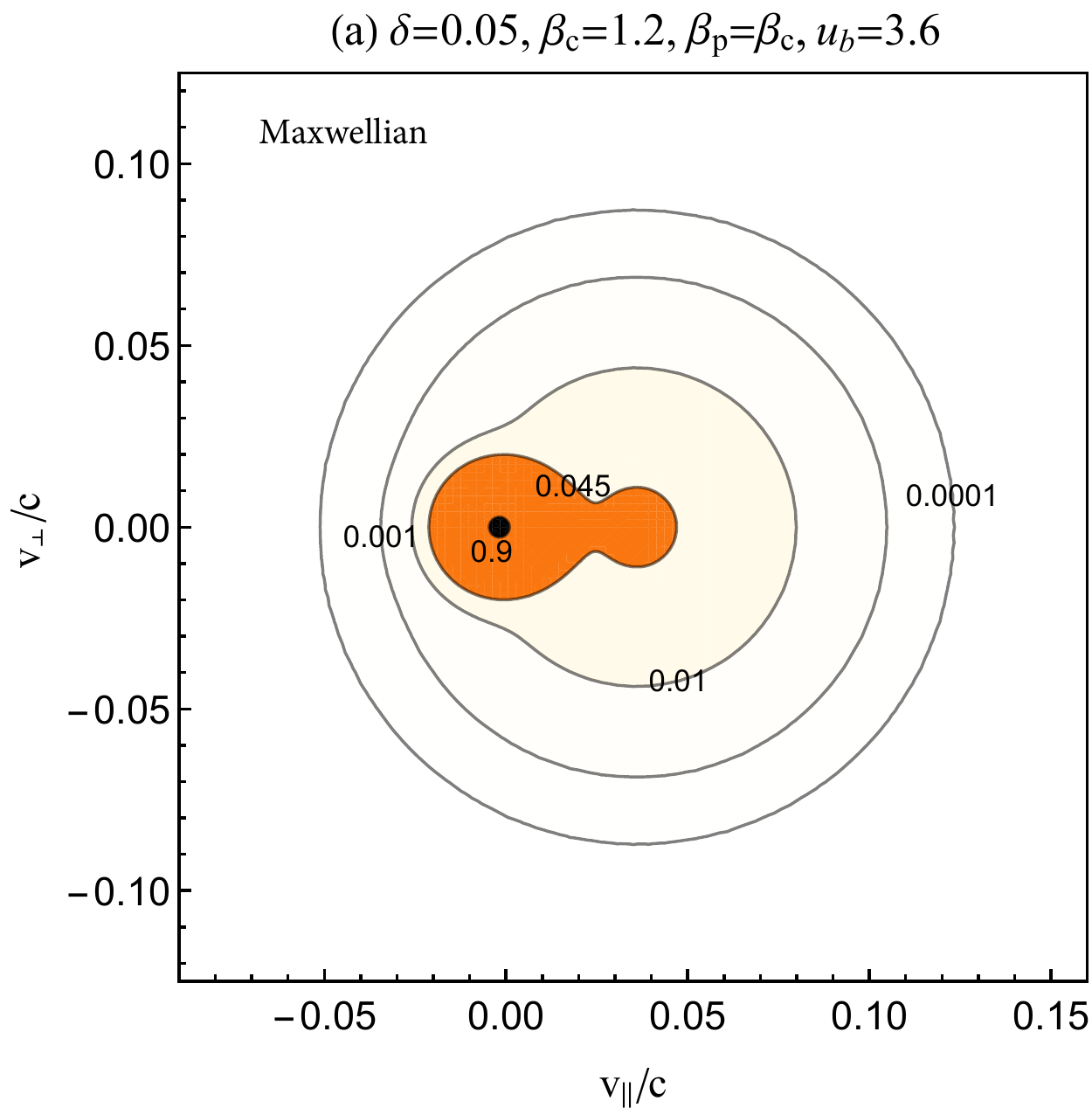}
\includegraphics[width=15.pc]{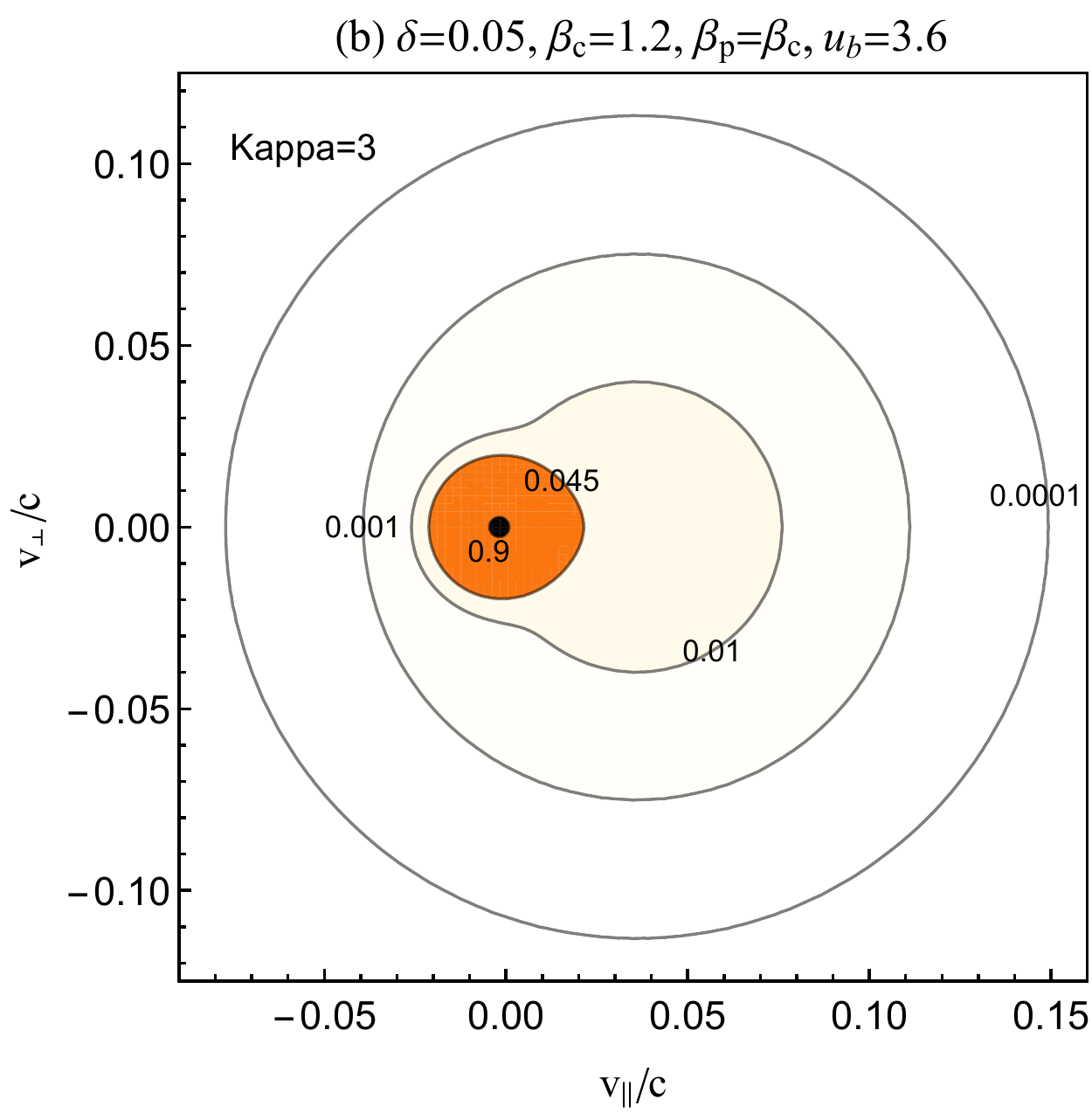}
\caption{Velocity distributions susceptible to FHF instability in Fig.\ref{fig:8} with a
Maxwellian ($\kappa\rightarrow\infty$) beam in panel (a), and a Kappa ($\kappa=3$) beam in
panel (b).}
\label{fig:9}
\end{figure}

In the presence of suprathermals the electron beam is well described by a drifting-Kappa 
distribution function. Fig.~\ref{fig:8} displays growth rates (panel a) and wave-frequency 
(panel b) of FHF instability and their variation with $\kappa$. Plasma parameters are the same 
as in Fig.~\ref{fig:7}, with $u_b=3.6$. Panel~(a) clearly shows that suprathermal electrons in the beam 
have an inhibiting effect on the FHF instability, i.e., the growth rates and the range of 
unstable wave numbers decrease with decreasing the power-index $\kappa$. The corresponding 
wave frequencies in panel~(b) are enhanced in the presence of suprathermal electrons, 
i.e., decreasing $\kappa$. These effects have not been reported by \cite{Saeed2017}, who
restricted to low beta regimes, and to Kappa approaches with $\kappa$-independent temperatures. 
Here we find that FHF instability is inhibited by the suprathermal electrons, by contrast to whistlers
which are stimulated by the same suprathermals. In Figure~\ref{fig:9}
we show explicitly the VDFs invoked to derive the unstable solutions in Fig.~\ref{fig:8}, for 
a Maxwellian beam ($\kappa \to \infty$) in panel~(a), and a Kappa beam ($\kappa=3$) in 
panel~(b) ($\beta_c=\beta_p=1.2$ and $\beta_b=12$). The beam-core contrast is diminished
in the presence of suprathermals (compare the orange contour in panels (a) and (b)), which 
may explain the inhibition of FHF instability and the stimulation of whistlers, as also 
shown by the velocity thresholds derived in the next section.  

\begin{figure}
\centering 
\includegraphics[width=19.5pc, trim={0 0.05cm 0 0},clip]{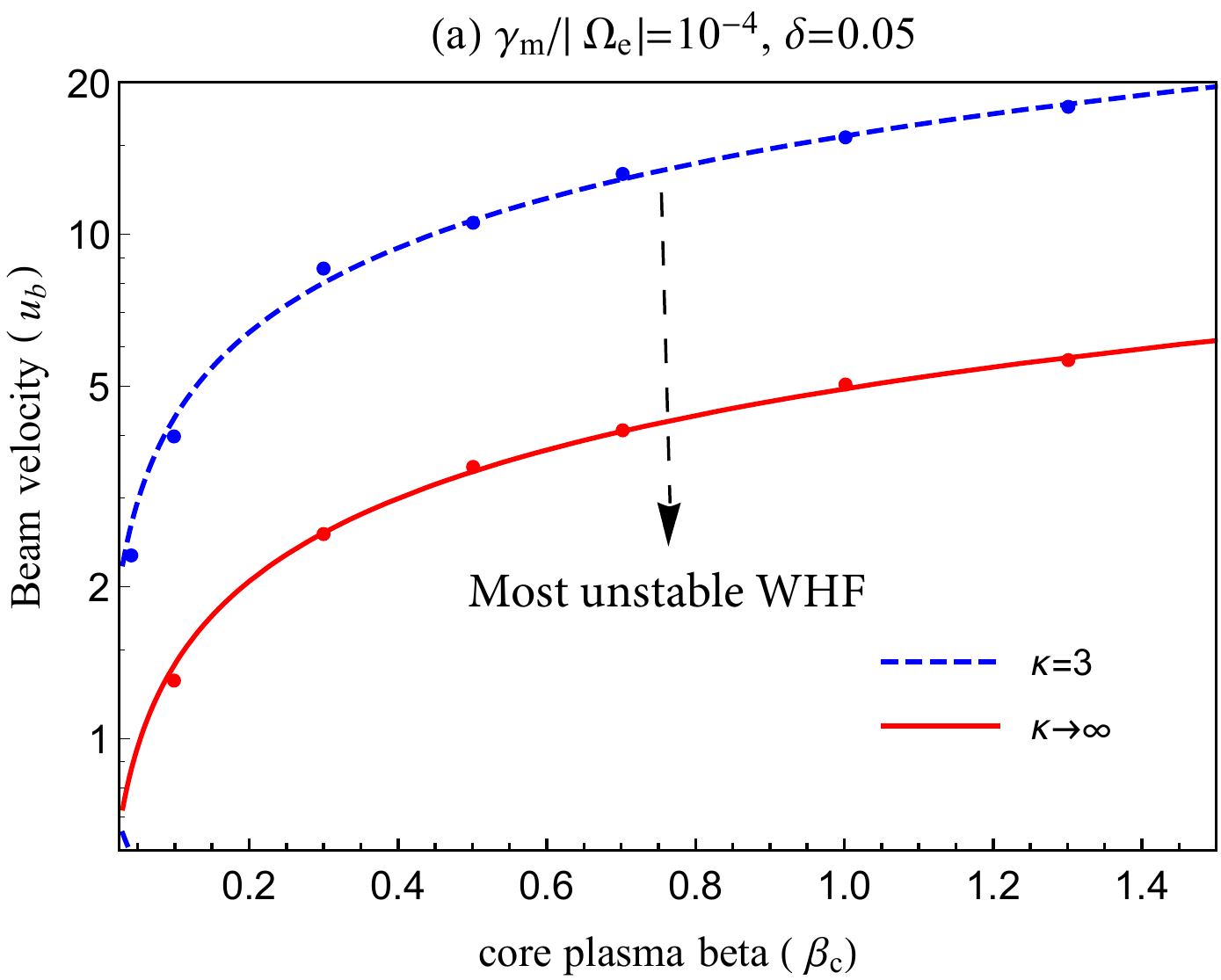}
\includegraphics[width=19.5pc, trim={0.05cm 0 0 0},clip]{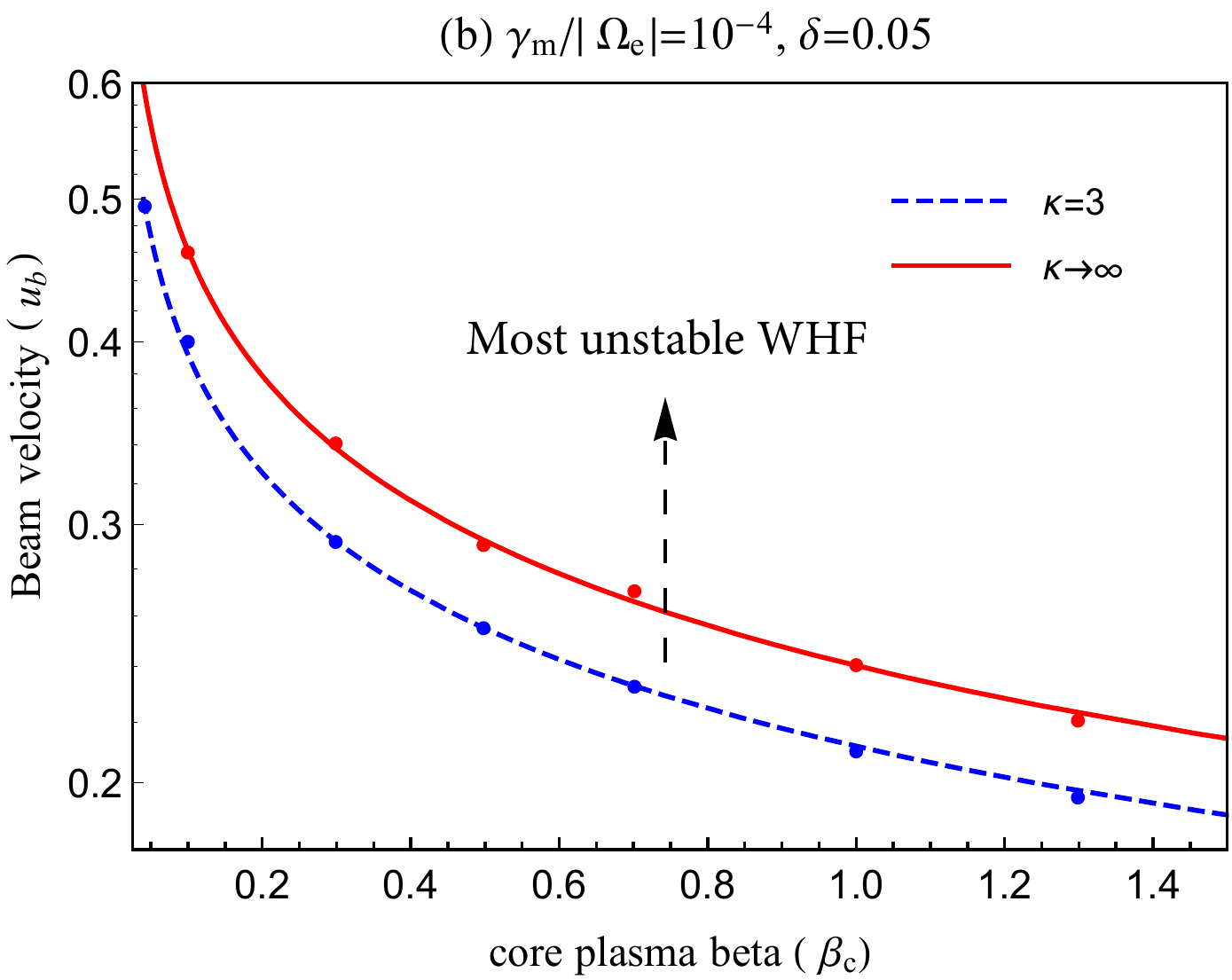}
\caption{The beam velocity upper (a) and lower (b) thresholds for the WHF instability 
(maximum growth rate $\gamma_m/|\Omega_e|=10^{-4}$) driven by Maxwellian (red) and Kappa
(blue) distributed beams. }
\label{fig:10}
\end{figure}
%
\begin{figure}
\centering 
\includegraphics[width=20pc]{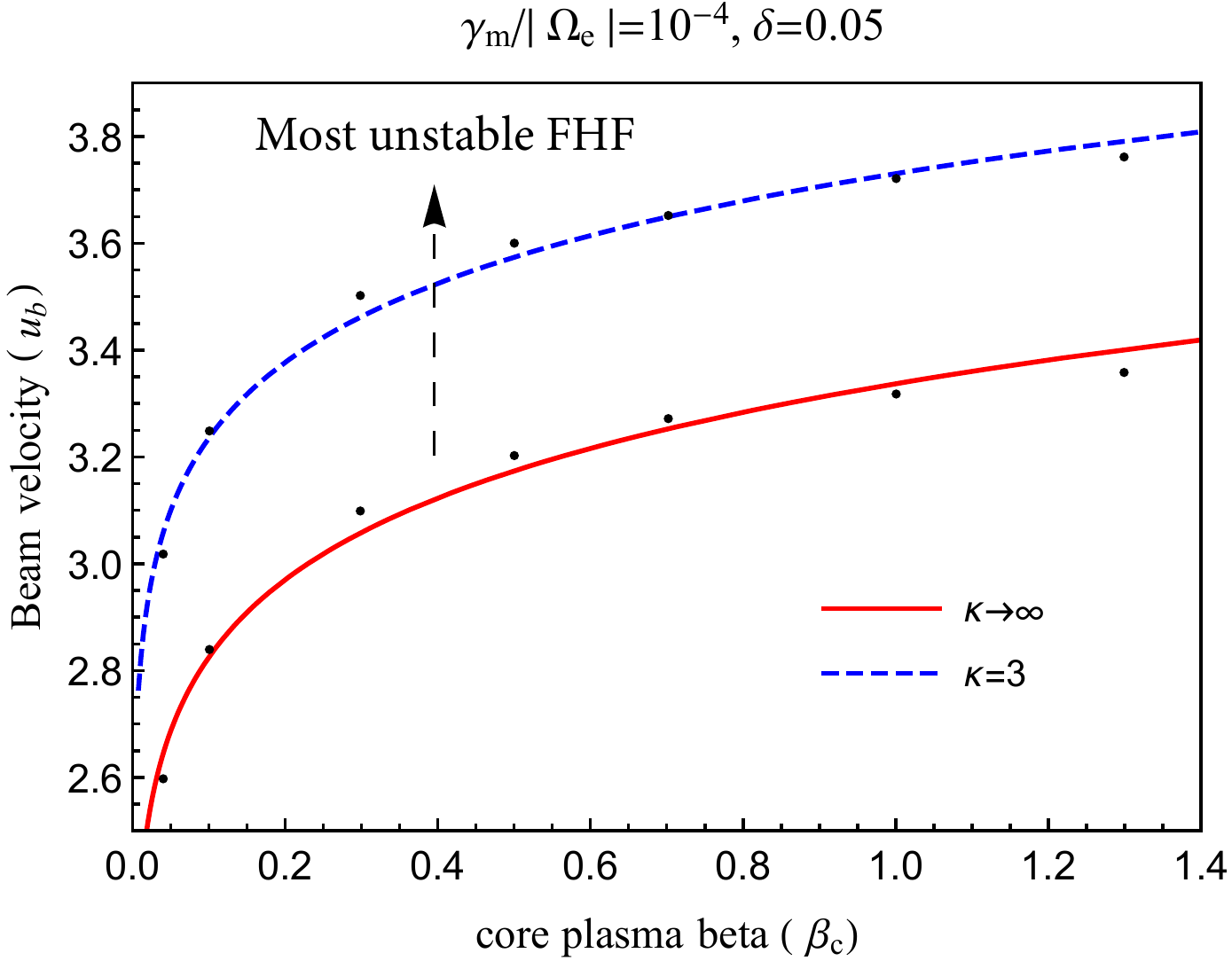}
\caption{The beam velocity threshold for the FHF instability ($\gamma_m/|\Omega_e|=10^{-4}$)
 driven by Maxwellian (red) and Kappa (blue) distributed beams. }
\label{fig:11}
\end{figure}

\section{Instability thresholds}

\begin{figure}
\centering 
\includegraphics[width=20pc]{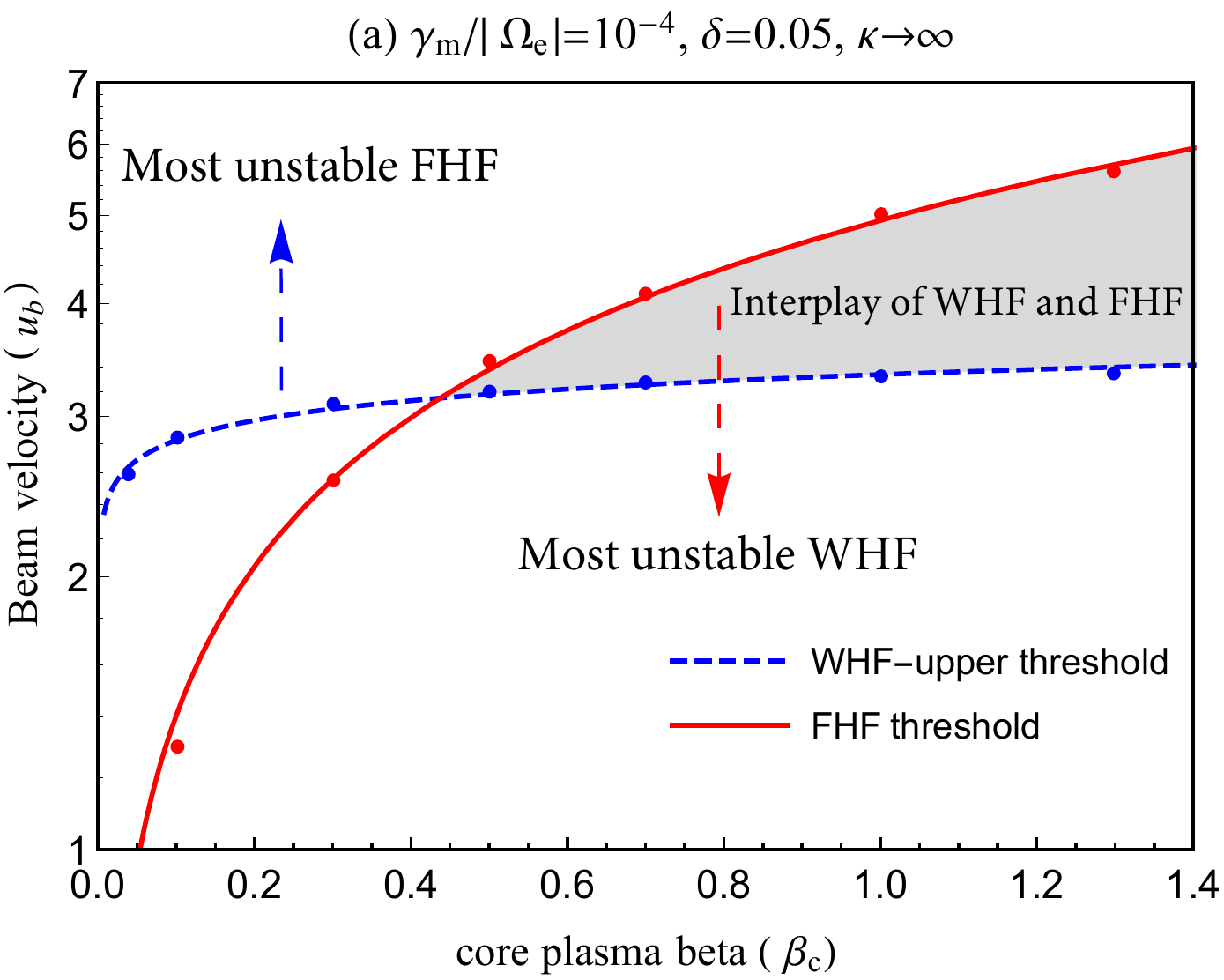}
\includegraphics[width=20pc]{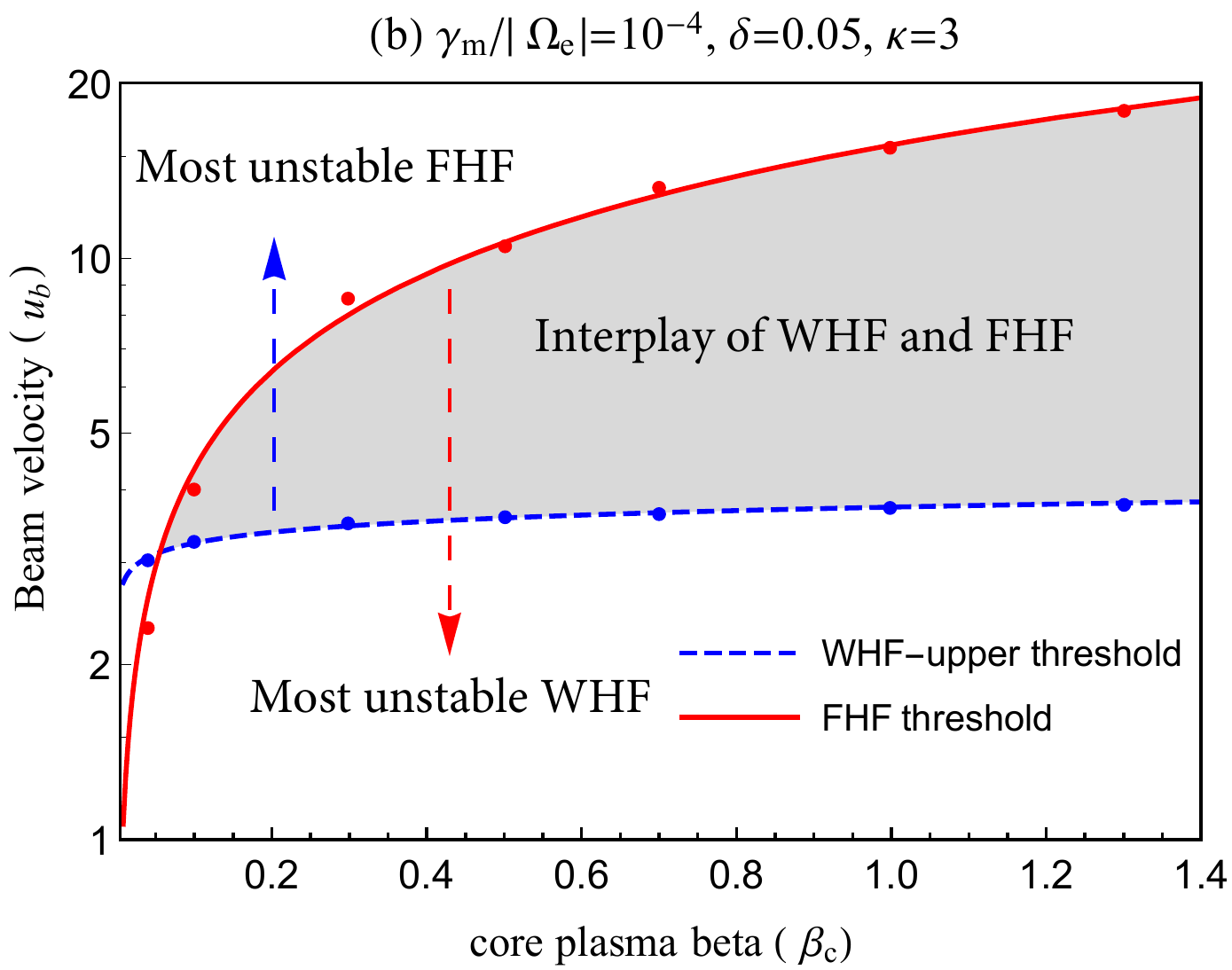}
\caption{Comparison of the instability thresholds ($\gamma_m/|\Omega_e|=10^{-4}$): WHF (red) vs. 
FHF (blue) for (a) Maxwellian and (b) Kappa beams.} \label{fig:12}
\end{figure}

Figs.~\ref{fig:10}--\ref{fig:12} present the beam velocity thresholds, which allow us to identify 
the regime of dominance of each instability, and the transitive conditions triggering the mode 
conversion. We know already that WHF instability is limited to low beam velocities (non-uniform 
variation of growth rates with $u_b$), and it is therefore expected to develop between two, 
upper and lower thresholds which are displayed in Figs.~\ref{fig:10}, panels~(a) and (b), 
respectively. These thresholds are derived for a maximum growth rate $\gamma_m= 10^{-4}|\Omega_e|$ 
approaching the marginal stability ($\gamma_m =0$), in terms of the main plasma parameters 
conditioning the instability, the beam velocity ($u_b)$ and the core plasma beta ($\beta_c$). 
Contours are obtained with an inverse power-law 
\begin{align}
u_{b}=\dfrac{a}{\beta_c^{~b}}, \label{e10}
\end{align} 
where $a$ and $b$ are the fitting parameters tabulated in Table~\ref{t2}. The most unstable regimes 
for the WHF modes are indicated by the dashed arrows, and are found between the upper and lower 
thresholds. 

%
\begin{table}
\centering
\caption{Fitting parameters in Figs.~\ref{fig:10} and \ref{fig:11}}
\label{t2}
\begin{tabular}{ccccccccccc}
\hline 
 & \multicolumn{2}{l}{Whistler (a)} & \multicolumn{2}{l}{Whistler (b)} & \multicolumn{2}{l}{Electron firehose} \\
                   &$a$  &      $b$     &  $a$   &    $b$ &  $a$ &$b$\\
 \hline 
$\infty$    & 4.94 &$-$0.545 & 0.240 & 0.283 &3.73 & $-$0.062  \\ 
 3               & 15.6 & $-$0.557 & 0.212& 0.266 & 3.34&$-$0.073  \\
 \hline 
\end{tabular}
\end{table}
%

The effects of of suprathermal beaming electrons are shown by contrasting thresholds obtained
for a Maxwellian beam (red contours) with those for a Kappa distributed beam with $\kappa = 3$ 
(blue contours). 
In panel~(a) the increase of the upper threshold with the core plasma beta means an extent of the
conditions favorable to WHF instability to higher values of beam velocity. Moreover, the upper 
threshold is markedly enhanced in the presence of suprathermal electrons, i.e., for $\kappa=3$, 
farther extending the instability conditions to even more energetic beams. Contrary to 
the upper thresholds, the lower thresholds in panel (b) decrease as the core beta increases, 
confirming the results in Fig.\ref{fig:2}. Again, the WHF instability is stimulated by the 
suprathermal electrons (for $\kappa =3$, confirming the results in Fig.~\ref{fig:6}), in this 
case by decreasing the beam velocity thresholds but increasing susceptibility to this instability.  


Fig.~\ref{fig:11} presents the beam velocity thresholds derived for the FHF instability ($\gamma_m=10^{-4}
|\Omega_e|$) in terms of the core plasma beta $\beta_c$, and for the same plasma parameters as in 
Fig.~\ref{fig:10}. Contours are obtained by fitting to the same law in Eq.~\eqref{e10} 
(see also Table~\ref{t2}). As explained before, the results in Fig~\ref{fig:5} suggest that for 
a higher thermal velocity (implying a higher $\beta_c$) we need more energetic beams, i.e., a higher
$u_b$, to excite the FHF instability. Variation with the core plasma beta resembles that of the 
upper threshold of WHF instability, but the most unstable FHF regimes are located above the FHF 
thresholds, as indicated by the dashed arrows in Fig.~\ref{fig:11}. Moreover, these thresholds 
increase as the power-index $\kappa$ decreases, confirming the inhibiting effect of the suprathermal 
beaming electrons on the FHF instability, described already in Fig.~\ref{fig:8}. 

Now a comparison of the instability thresholds becomes straightforward, and enables to build 
a clear picture on the interplay of these instabilities and their regimes of dominance. 
As illustrated in Fig.\ref{fig:11} the most unstable FHF modes are obtained for beams with relatively 
high velocities $u_b > 2.5$, making relevant only a contrast with the upper threshold of WHF instability.
This contrast is provided in Fig.\ref{fig:12} for a Maxwellian beam ($\kappa\rightarrow \infty$) in 
panel (a) and for a Kappa beam ($\kappa=3$) in panel~(b). Dashed arrows in Figure~\ref{fig:12} 
indicate the most unstable regime for each instability.
Panel (a) shows clearly that the WHF instability is dominant at low beaming velocities, e.g., $u_b<2.5$, 
while for higher beam velocities the FHF instability arises and eventually dominates, if the 
core plasma beta is low enough, e.g., $\beta_c<0.5$, or if the beam is energetic enough, with 
high $u_b$ exceeding the threshold of WHF. If $\beta_c > 0.5$ is large enough we identify a 
regime where both WHF and FHF instabilities may coexist, i.e., gray area between their thresholds. 
We have already shown a representative case for the interplay of these two instabilities in 
Fig.\ref{fig:4}. The interval of beam velocities relevant for this regime of transition and 
interplay increases with the core plasma beta. This regime further expands in the presence 
of the suprathermal beaming electrons, see panel (b) ($\kappa=3$), when the limit value of the 
core plasma beta becomes also significantly lower, $\beta_c>0.05$.

\section{Conclusive discussions} 

In the present paper, we have clarified the main contrasting properties of heat-flux instabilities, namely 
the WHF and FHF instabilities, driven by the electron beam-core relative drift. Thermal and 
suprathermal spread of electrons plays a key role, and here we have assumed isotropic 
temperatures, focusing on the effect of beaming electrons as a unique source of free energy. 
In an attempt to overcome the limitations from previous studies and reproduce conditions 
typically encountered in space plasmas, here we have considered an extended range of values 
for the key plasma parameters, e.g., the beam velocity and the core plasma beta, and for the 
electron beam a more realistic (drifting-)Kappa distribution enabling a direct contrast to 
the Maxwellian limit of lower temperature. 

Less energetic beams destabilize the whistler modes in the wave-frequency range $\Omega_p
\ll \omega_r \ll |\Omega_e|$. This is the WHF instability with growth-rates conditioned by the thermal 
velocity of the resonant beam (Figs.~\ref{fig:2} and \ref{fig:3}), and non-uniformly varying with 
the beam velocity (Fig.~\ref{fig:1}). The fastest growing modes are described by the maximum growth 
rates derived in Fig.~\ref{fig:2} in terms of beam velocity threshold $u_{bt}$ and the core plasma beta 
parameter $\beta_c$. Growth rates are maximized by increasing any of these two parameters.
Beaming velocities higher than this threshold can excite the LH-FHF growth-rates, which display an 
additional peak at lower wave-numbers, see Fig.~\ref{fig:4}. A qualitative explanation for this transitive  
regime is provided by the contours of the distributions in Figs.~\ref{fig:3} and \ref{fig:5}, which 
show that beaming electrons become less resonant for higher beaming velocities, inhibiting the WHF modes
and exciting the FHF instability. These results confirm the earlier predictions in \cite{Gary1985}.
The FHF instability has an opposite LH polarization and is excited for higher beaming velocities.
This unstable mode may evolve out of a RH whistler mode for significant beaming velocities 
and/or core plasma beta parameters. Further increase of the beam velocity leads to a uniform 
increase of growth rates and wave frequencies. These transitions and conversions of the wave 
polarization are presented in Figs.~\ref{fig:7} and \ref{fig:8}.

Electron heat flux is transported away from the solar corona mainly by the suprathermal strahl
(or beaming) component with a drifting-Kappa distribution. According to our knowledge, in the 
existing studies of heat flux instabilities the effects of suprathermals are underestimated
by assuming Maxwellian beams or Kappa beams of comparable (kinetic) temperature, e.g., in 
\cite{Saeed2017}. However, in a realistic Kappa approach as the one invoked here, suprathermal 
electrons contribute with an excess of kinetic (free) energy \citep{Lazar2015, Lazar2017a} that
stimulates WHF instability by enhancing the growth rates and extending the instability regime to 
higher beaming velocities, e.g., in Fig.~\ref{fig:6}. For more energetic beams susceptible to 
FHF instability, the same suprathermal electrons have an opposite effect, diminishing the growth 
rates and the range of unstable wave-numbers. Physical explanations are suggested by the contours 
of the distributions in Fig.~\ref{fig:9}, which show that the beam-core contrast, and, implicitly,
the effective anisotropy in the parallel direction are reduced in the presence of suprathermal 
electrons, causing inhibition of FHF modes and stimulation of WHF instability.

These physical insights have enabled us to identify the regimes of dominance for each of these
two instabilities, in terms of the instability thresholds derived in Figs.~\ref{fig:10}--\ref{fig:11}. 
The most unstable WHF modes are located between two thresholds, 
namely, a lower and an upper threshold in Fig.~\ref{fig:10}, while the most unstable FHF modes 
are located upper the threshold in Fig.~\ref{fig:11}. As a consequence of that, with increasing
 the core plasma beta, conditions favorable to WHF instability are enhanced while those favorable
to FHF are reduced. These effects are markedly stimulated in the presence of the suprathermal 
beaming electrons. For a low beam velocity $u_b \leqslant 2.5$ the WHF instability is dominant 
and the FHF modes are damped, which totally agree with the results for FHF instability in 
\cite{Saeed2017b}. However, for more energetic beams, we have identified the regime of transition
and mode conversion, where both instabilities can develop and compete in the relaxation process. 
Marginally bounded by the FHF threshold and the upper WHF threshold, the range of beaming 
velocities associated to this regime is considerably enhanced with increasing $\beta_c$, and 
due to the abundance of suprathermal electrons (i.e., lowering $\kappa$). 

To conclude, in this paper we have decoded the interplay of HF instabilities conditioned 
by the relative beaming velocity of two countermoving electron populations, and, in particular, 
we have unveiled new unstable regimes induced or/and stimulated by the suprathermal beaming 
electrons. Suprathermal Kappa-distributed electrons are ubiquitous in the solar wind and their 
effects support earlier predictions \citep{Gary1985, Gary2000} which indicate the WHF instability 
as  the most probable mechanism of regularization of the electron strahl. However,
an extended analysis to include oblique modes, like the aperiodic branch of firehose instability,
may also be opportune, especially for energetic electron beams. Our present results do offer 
valuable tools for future investigations seeking realistic approach of the electron
beams and their implications in specific conditions, e.g., fast winds and interplanetary shocks.
These conditions should also include the effect of temperature anisotropy of electrons, 
which is often reported by the solar wind observations \citep{Stverak2008} and, eventually,
the reaction of protons to the low-frequency fluctuations developed by the firehose-like
instabilities \citep{Sarfraz2017, Yoon2017}.

\section*{Acknowledgements}

These results were obtained in the framework of the projects SCHL 201/35-1 (DFG-German 
Research Foundation), GOA/2015-014 (KU Leuven), G0A2316N (FWO-Vlaanderen), and C 90347 
(ESA Prodex 9). S.M. Shaaban acknowledges support by a FWO Postdoctoral Fellowship (Grant No. 12Z6218N), and a FWO Travel grant for long stay aboard (Grant No. V419818N). The authors acknowledge useful discussions during the first meeting of the ISSI team "Kappa Distributions: From Consistent Theory of Suprathermal Space Plasmas".




\bibliographystyle{mnras}
\bibliography{papers} 


\bsp	
\label{lastpage}
\end{document}